\documentclass[12pt,english, journal,draftclsnofoot,onecolumn]{IEEEtran}
\usepackage[T1]{fontenc}
\usepackage[latin9]{inputenc}
\usepackage{geometry}
\usepackage{float}
\usepackage{textcomp}
\usepackage{mathrsfs}
\usepackage{amsthm}
\usepackage{amsmath}
\usepackage{stackrel}
\usepackage{graphicx}
\usepackage{setspace}
\makeatletter
\theoremstyle{plain}
\newtheorem{thm}{\protect\theoremname}
\theoremstyle{plain}
\newtheorem{lem}[thm]{\protect\lemmaname}

\@ifundefined{showcaptionsetup}{}{%
 \PassOptionsToPackage{caption=false}{subfig}}
\usepackage{subfig}
\makeatother

\usepackage{babel}
\providecommand{\lemmaname}{Lemma}
\providecommand{\theoremname}{Theorem}

\begin{document}

\title{Impact of Phase-Shift Error on the Secrecy Performance of Uplink
RIS Communication Systems}

\author{Abdelhamid Salem,\textit{\normalsize{} Member, IEEE}{\normalsize{},}
Kai-Kit Wong, \textit{\normalsize{}Fellow, IEEE}, and Chan-Byoung
Chae,\textit{\normalsize{} Fellow, IEEE}\\
\thanks{Abdelhamid Salem is with the department of Electronic and Electrical
Engineering, University College London, London, UK, (emails: a.salem@ucl.ac.uk).

Kai-Kit Wong is with the department of Electronic and Electrical Engineering,
University College London, London, UK, Kai-Kit Wong is also affiliated
with Yonsei University, Seoul, Korea (email: kai-kit.wong@ucl.ac.uk). 

Chan-Byoung Chae is with Yonsei University, Seoul, Korea (e-mail:
cbchae@yonsei.ac.kr). 

The work is supported by the Engineering and Physical Sciences Research
Council (EPSRC) under grant EP/V052942/1. For the purpose of open
access, the authors will apply a Creative Commons Attribution (CCBY)
licence to any Author Accepted Manuscript version arising.%
} }
\maketitle
\begin{abstract}
Reconfigurable intelligent surface (RIS) has been recognized as a
promising technique for the sixth generation (6G) of mobile communication
networks. The key feature of RIS is to reconfigure the propagation
environment via smart signal reflections. In addition, active RIS
schemes have been recently proposed to overcome the deep path loss
attenuation inherent in the RIS-aided communication systems. Accordingly,
this paper considers the secrecy performance of up-link RIS-aided
multiple users multiple-input single-output (MU-MISO) communication
systems, in the presence of multiple passive eavesdroppers. In contrast
to the existing works, we investigate the impact of the RIS phase
shift errors on the secrecy performance. Taking into account the complex
environment, where a general Rician channel model is adopted for all
the communication links, closed-form approximate expressions for the
ergodic secrecy rate are derived for three RIS configurations, namely,
i) passive RIS, ii) active RIS, iii) active RIS with energy harvesting
(EH RIS). Then, based on the derived expressions, we optimize the
phase shifts at the RIS to enhance the system performance. In addition,
the best RIS configuration selection is considered for a given target
secrecy rate and amount of the power available at the users. Finally,
Monte-Carlo simulations are provided to verify the accuracy of the
analysis, and the impact of different system parameters on the secrecy
performance is investigated. The results in this paper show that,
an active RIS scheme can be implemented to enhance the secrecy performance
of RIS-aided communication systems with phase shift errors, especially
when the users have limited transmission power.\end{abstract}


\section{Introduction}

Reconfigurable intelligent surface (RIS), also known as intelligent
reflecting surface (IRS), has been proposed recently as a promising
technique to extend the coverage and improve the spectral efficiency
of wireless communication networks \cite{ref1,Reef1}. Specifically,
RIS is composed of reflecting elements, each of which independently
imposes a phase shift on the incident signals. By tuning the phase
shifts of the reflecting elements, RIS can convert the propagation
environments into smart ones and thus enhance the received signals
quality \cite{ref1,Reef1}. Due to these advantages, RIS techniques
have been extensively considered in the literature. For instance,
in \cite{Ref2}, the fundamental capacity limit of RIS-aided multiple-input
multipleoutput (MIMO) communication systems has been considered. The
achievable ergodic rate of a RIS-assisted MIMO system which comprises
links of a Rician channel was derived in \cite{Ref3}. In \cite{Ref4},
a closed-form asymptotic ergodic sum rate of a RIS-assisted MIMO communication
system was derived under the assumption that the number of base station
(BS) antennas tends to infinity. In \cite{Ref5}, the up-link achievable
rate in RIS-aided massive MIMO systems has been analyzed and optimized.
The authors in \cite{Ref6,Ref7} analyzed the achievable rate of RIS-assisted
multiple users (MU) up-link massive MIMO system under Rician fading
channels. In \cite{Ref8,Ref9}, a closed-form expression of ergodic
achievable rate for RIS-aided massive MIMO systems with zero forcing
(ZF) detector has been derived. In addition, a closed-form analytical
expression for the symbol error probability and the upper bound on
the channel capacity of a RIS communication system have been derived
in \cite{Ref10}. The work in \cite{Ref11} considered the impact
of hardware impairments on a general RIS MU-MISO system with Rayleigh
fading channels. The ergodic capacity of RIS MIMO networks over Rayleigh-Rician
channels was considered in \cite{Ref12}. 

However, the practical implementation of passive RIS-aided communication
systems may face several challenges. For instance, the transmitted
signal propagates through the RIS experiences a double-fading attenuation,
e.g, source-RIS and RIS-destination links. This issue has been tackled
in the literature by increasing the number of passive RIS elements
\cite{Ref13}. However, this solution leads to an increase in the
size of the RIS module, which is impractical in some scenarios. To
tackle this issue the authors in \cite{Ref14} proposed RIS with active
elements. The main idea of active RIS is to adjust the phase shifts
and also amplify the reflected signal attenuated from the first link
with extra power consumption. Theoretical comparison between the active
RIS-assisted system and the passive RIS-aided system has been presented
in \cite{Ref15}. The results in \cite{Ref15} show that the active
RIS has better performance than passive RIS. The use of active RIS
elements to overcome the double-fading problem has been also investigated
in \cite{Ref16}, where the results illustrated that using active
elements results in a severe reduction in the physical size of RIS
to achieve a certain performance. To reduce the power consumption
of active RIS, a sub-connected architecture has been proposed in \cite{Ref17}.
The energy efficiency in an active RIS-aided MU-MISO down-link system
has been investigated in \cite{Ref18}. 

Although fixed embedded batteries can be used to power the RIS, these
batteries cannot be relied on for long time and uninterrupted operations.
In addition, wired charging might not be possible to use if the RIS
is deployed in inaccessible places. Therefore, equipping RIS elements
with energy harvesting (EH) modules can solve these issues. Accordingly,
a self-sustainable RIS approach was proposed and studied in the resent
researches on RIS. In this regard, in \cite{Ref19} time switching
(TS) and power splitting (PS) EH protocols for the RIS to harvest
sufficient amount of energy from an access point have been proposed
and investigated. The work in \cite{Ref20} considered a self-sustainable
RIS-aided MU-MISO communication systems, in which the RIS collected
energy from the radio frequency (RF) transmitter using the PS protocol.
In \cite{Ref21}, a novel transmission policy for a communication
network assisted by self-sustainable RIS has been proposed, where
the RIS harvests energy from an energy transmitter to support its
operation. In \cite{Ref22}, self-sustainable RIS with the PS protocol
to assist broadcasting network was studied. In \cite{Ref23}, self-sustainable
RIS-aided communication between a gateway and a device was studied,
in which the RIS harvested energy prior communication.

Moreover, due to the broadcast nature of wireless channels, confidential
messages are vulnerable to eavesdropping attacks. For the provision
of secure transmission, physical layer security (PHYSec) has been
proposed from the information theory perspective \cite{Ref24,Ref25}.
PHYSec exploits the nature of wireless channels to enhance the system
security \cite{Ref24,Ref25}. PHYSec of RIS systems has also been
studied in the literature. In \cite{Ref26}, the secrecy throughput
maximization problem has been formulated and solved to enhance the
secrecy performance of the RIS-assisted MIMO systems. In \cite{Ref27},
a novel active RIS design to enhance the security of wireless transmission
was proposed. PHYSec of RIS-aided wireless networks has been considered
in \cite{Ref28} to achieve secure transmission between a source and
a legitimate user in the presence of a malicious eavesdropper. In
\cite{Ref29}, RIS has been used to perform secure transmission from
a multiple antennas transmitter to a multiple antennas legitimate
receiver. Further work in \cite{Ref30} considered the secrecy transmission
in a RIS-aided multiple antennas communication, where the secrecy
rate was improved by optimizing the RIS location. In \cite{Ref31},
an active RIS-aided multiple antennas PHYSec transmission scheme was
considered, where the active RIS was designed to amplify the signal
actively.

Accordingly, this paper investigates the impact of phase shift error
on the secrecy performance of up-link RIS-aided MU-MISO systems in
the presence of multiple eavesdroppers. The BS receives the users
messages only through the RIS, while eavesdroppers can receive the
signals from both the direct and reflected links. Under Rician fading
channels and phase shift errors, the ergodic secrecy rate is analyzed
for three RIS configurations, namely, 1) passive RIS, 2) active RIS,
and 3) EH RIS. Based on the derived rate expressions, the phase shifts
at the RIS are optimized to enhance the system performance. Then,
the best RIS configuration selection is considered based on the target
secrecy rate and amount of power available at the users. For clarity
we list the main contributions of this work as follows: 

1) We investigate the impact of RIS phase shift error on the secrecy
performance of up-link MU-MIMO systems in the presence of multiple
passive eavesdroppers.

2) New closed-form explicit analytical expressions for the ergodic
secrecy rate are derived for the RIS-assisted MU-MIMO systems, when
the RIS is passive, active and EH node under Rician fading channels.
This channel model is more general but also very challenging to be
considered mathematically. The derived secrecy rate expressions are
simple, explicit and in closed form, and provide several important
practical design insights. 

3) Based on the derived expressions, a genetic algorithm (GA)-based
approach is used to obtain the optimal phase shifts. Also, a simple
suboptimal technique is proposed to enhance the secrecy rate for a
legitimate user. 

4) Given a target secrecy rate, we calculate the required user power,
and we present steps to select best RIS configuration which depend
mainly on the available power at the users.

5) Finally, Monte-Carlo simulations are performed to validate the
analytical expressions. Then, the impact of several system parameters
on the secrecy performance are investigated. 

The results in this work show that active RIS is an efficient scheme
to achieve secure communication in the presence of phase shift errors
at the RIS, especially when there is no sufficient amount of power
at the users. 

Next, Section \ref{sec:System-Model} presents the RIS-aided uplink
MU-MISO system model. In Section \ref{sec:Passive-RIS}, we derive
the ergodic secrecy rate of the passive RIS model. Section \ref{sec:Active-RIS}
presents the ergodic secrecy rate of the active RIS scheme. Section
\ref{sec:EH-RIS} derives the ergodic secrecy rate of the EH RIS scheme.
Section \ref{sec:Numerical-Results} depicts our numerical results.
Our main conclusions are summarized in Section \ref{sec:Conclusions}.

\section{System Model\label{sec:System-Model}}

Consider a typical up-link RIS-aided MU-MISO communication system
consisting of a multiple antennas BS, an RIS and $K$ single-antenna
users in the presence of $J$ single antenna passive eavesdroppers.
The BS is equipped with $N$ antennas, and the RIS is equipped with
$M$ reflecting elements, as shown in Fig. 1. 
\begin{figure}[H]
\noindent \begin{centering}
\includegraphics[bb=230bp 150bp 680bp 500bp,clip,scale=0.4]{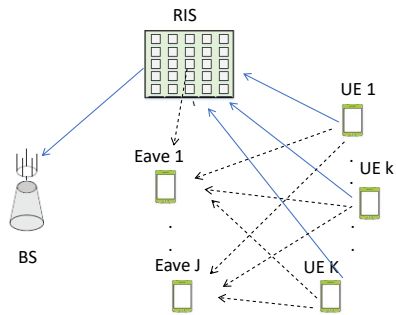}
\par\end{centering}

\protect\caption{An RIS-aided uplink MU-MISO system with $N$ BS antennas, $M$ RIS
elements, $K$ users and $J$ eavesdroppers.}
\end{figure}
 The BS and RIS are connected to control and adjust the phase shifts
of the the RIS elements. It is assumed that the eavesdroppers can
hear the signals from the direct and reflected links, and trying to
eavesdrop a specific confidential message in the system. On the other
side, the direct links between the users and BS are assumed to be
blocked, which justifies the use of the RIS. It is known that, the
RIS is most likely to be installed on the buildings, and thus it can
create channels dominated by line-of-sight (LoS) path along with scatters.
Accordingly, a Rician fading model is considered for the RIS channels.
The channel matrix between the RIS and the BS is denoted by $\mathbf{G}\in C^{N\times M}$
, and the channel vector between user $k$ and the RIS is presented
by $\mathbf{h}_{r,k}\in C^{M\times1}$. The mathematical expressions
of the channel matrix $\mathbf{G}$ and the channel vector $\mathbf{h}_{r,k}$
can be expressed, respectively, as

\begin{equation}
\mathbf{G}=\left(\sqrt{\frac{\rho_{b}}{\rho_{b}+1}}\mathbf{\bar{G}}+\sqrt{\frac{1}{\rho_{b}+1}}\mathbf{\tilde{G}}\right),\quad\mathbf{h}_{r,k}=\left(\sqrt{\frac{\rho_{k}}{\rho_{k}+1}}\mathbf{\bar{h}}_{r,k}+\sqrt{\frac{1}{\rho_{k}+1}}\mathbf{\tilde{h}}_{r,k}\right)
\end{equation}

\noindent where $\rho_{b}$ and $\rho_{k}$ are the Rician factors,
$\mathbf{\bar{G}}$ and $\mathbf{\bar{h}}_{r,k}$ are the LoS components
and $\mathbf{\tilde{G}}$ and $\mathbf{\tilde{h}}_{r,k}$ are the
NLoS components, in which

\begin{equation}
\mathbf{\bar{G}}=\mathbf{a}_{N}\left(\phi_{r}^{a},\phi_{r}^{e}\right)\mathbf{a}_{M}^{H}\left(\phi_{t}^{a},\phi_{t}^{e}\right),\quad\mathbf{\bar{h}}_{r,k}=\mathbf{a}_{M}\left(\phi_{kr}^{a},\phi_{kr}^{e}\right)
\end{equation}

\noindent where $\phi_{kr}^{a},\phi_{kr}^{e}$ denote the azimuth
and elevation angles of arrival (AoA) from user $k$ to the RIS ,
respectively, $\phi_{t}^{a},\phi_{t}^{e}$ are the azimuth and elevation
angles of departure (AoD) at the BS from the RIS, respectively, $\phi_{r}^{a},\phi_{r}^{e}$
are the azimuth and elevation AoA from the RIS to the BS, respectively.
The $k$th element of the vector \textbf{$\mathbf{a}_{X}$} can be
written as $\left[\mathbf{a}_{X}\left(\phi_{1},\phi_{2}\right)\right]_{k}=e^{j2\pi\frac{d}{\lambda}\left(x_{k}\sin\phi_{1}\sin\phi_{2}+y_{k}\cos\phi_{2}\right)},$
where $\lambda$ is the wavelength, $d$ is the elements/antennas
spacing, and $x_{k}=\left(k-1\right)\textrm{mod}\sqrt{x}$, $y_{k}=\frac{k-1}{\sqrt{X}}$.
On the other hand, the channel vector between the RIS and eavesdropper
$j$ is presented by $\mathbf{h}_{e_{j},r}\in C^{1\times M}$ , and
the channel from user $k$ to eavesdropper $j$ is $h_{e_{j},k}\in C^{1\times1}$.
The direct channel fading is assumed to be Rayleigh fading due to
extensive scatterers, while for the RIS-related channels, is assumed
to be Rician fading. Thus the expression of $\mathbf{h}_{e_{j},r}$is
given by 

\begin{equation}
\mathbf{h}_{e_{j},r}=\left(\sqrt{\frac{\rho_{e_{j},r}}{\rho_{e_{j},r}+1}}\bar{\mathbf{h}}_{e_{j},r}+\sqrt{\frac{1}{\rho_{e_{j},r}+1}}\tilde{\mathbf{h}}_{e_{j},r}\right)
\end{equation}

\noindent where $\rho_{e_{j},r}$ is the Rician factor, $\bar{\mathbf{h}}_{e_{j},r}$
and $\tilde{\mathbf{h}}_{e_{j},r}$ are the LoS of NLoS components,
respectively.

The channel state information (CSI) of the eavesdroppers is assumed
to be unknown at the BS/RIS (only statistical information can be known),
and the eavesdroppers are non-colluding. Therefore, the ergodic secrecy
rate can be calculated by \cite{ergodicSR}

\begin{equation}
\hat{R}_{s}=\left[\hat{R}_{b_{k}}-\hat{R}_{e_{j,k}}\right]^{+}
\end{equation}

\noindent where $[l]^{+}$$=\textrm{max}\,\left(0,l\right)$, $\hat{R}_{b_{k}}=\mathscr{E}\left\{ R_{b_{k}}\right\} $,
$R_{b_{k}}$ is the up-link rate of user $k$, and $\hat{R}_{e_{j,k}}=\max\mathscr{E}\left\{ R_{e_{j,k}}\right\} $,
$R_{e_{j,k}}$ is the rate at eavesdropper $j$. 

In the following sections, we consider the secrecy performance of
the three RIS configurations.

\section{Passive RIS \label{sec:Passive-RIS}}

As we have mentioned earlier, passive RIS reflects the users messages
constructively to the BS with passive elements. Thus, the received
signal at the BS can be expressed as

\begin{equation}
\mathbf{y}_{b}=\stackrel[k=1]{K}{\sum}\sqrt{p_{k}\, L_{u_{k},b}}\mathbf{G}\tilde{\Theta}\mathbf{h}_{r,k}x_{k}+\mathbf{n}_{b}
\end{equation}

\noindent where $L_{u_{k},b}=d_{u_{k},r}^{-\alpha_{r}}d_{r,b}^{-\alpha_{b}}$
is the large scale fading, $d_{u_{k},r}$ is the distance between
user $k$ and RIS, $d_{r,b}$ is the distance between RIS and the
BS, $\alpha_{r}$ and $\alpha_{b}$ are the path-loss exponents, $\mathbf{n}_{b}$
is the additive wight Gaussian noise (AWGN) at the BS, $\mathbf{n}_{b}\sim CN\left(0,\sigma_{b}^{2}\mathbf{I}\right)$,
$\tilde{\Theta}=\bar{\Theta}\Theta$ where $\Theta=\textrm{diag}\left(\theta\right)$,
and $\theta=\left[\theta_{1},......,\theta_{M}\right]^{T}$is the
RIS reflection coefficients with $\theta_{m}=e^{j\varphi_{m}}$, where
$\varphi_{m}\text{\ensuremath{\in}}[0,2\pi)$ is the phase shift of
element $m$. However, in practical systems, phase shift errors can
exist due to imperfect channel knowledge and finite precision in phase
adjustment. Thus, we define $\bar{\Theta}=\left[e^{j\bar{\varphi_{1}}},....,e^{j\bar{\varphi_{M}}}\right]$
as the phase-shift errors at the RIS. The phase-error is modeled according
to Von-Mises (VM) distribution with zero-mean and a characteristic
function (CF) $E\left[e^{j\bar{\varphi_{m}}}\right]=\frac{I_{1}\left(\kappa\right)}{I_{0}\left(\kappa\right)}=\rho\left(\kappa\right)$,
where $\kappa$ is the concentration parameter and $I_{i}$ is the
modified Bessel function of the first kind and order $i$. By applying
the receive beamforming vector $\mathbf{w}_{k}$ at the BS, the received
signal of user $k$ is 

\begin{equation}
\mathbf{y}_{b,k}=\sqrt{p_{k}\, L_{u_{k},b}}\mathbf{w}_{k}\mathbf{G}\bar{\Theta}\Theta\mathbf{h}_{r,k}x_{k}+\stackrel[\underset{i\neq k}{i=1}]{K}{\sum}\sqrt{p_{i}\, L_{u_{i},b}}\mathbf{w}_{k}\mathbf{G}\bar{\Theta}\Theta\mathbf{h}_{r,i}x_{i}+\mathbf{w}_{k}\mathbf{n}_{b}.
\end{equation}

On the other hand, the received signal at eavesdropper $j$ to detect
user $k$ signal is

\[
y_{e_{j,k}}=\sqrt{p_{k}}x_{k}\left(\sqrt{d_{e_{j},k}^{-\alpha_{e}}}h_{e_{j},k}+\sqrt{L_{u_{k},e_{j}}}\mathbf{h}_{e_{j},r}\bar{\Theta}\Theta\mathbf{h}_{r,k}\right)
\]

\begin{equation}
+\stackrel[\underset{i\neq k}{i=1}]{K}{\sum}\sqrt{p_{i}}x_{i}\left(\sqrt{\, d_{e_{j},i}^{-\alpha_{e}}}h_{e_{j},i}+\stackrel[\underset{i\neq k}{i=1}]{K}{\sum}\sqrt{L_{u_{i},e_{j}}}\mathbf{h}_{e_{j},r}\bar{\Theta}\Theta\mathbf{h}_{r,i}\right)+n_{e_{j}}
\end{equation}

\noindent where $d_{e_{j},k}^{-\alpha_{e}}$ is the distance between
user $k$ and eavesdropper $j$, $\alpha_{e}$ is the path-loss exponent,
$L_{u_{k},B}=d_{u_{k},r}^{-\alpha_{r}}d_{e_{j},r}^{-e}$ and $d_{e_{j},r}^{-e}$
denotes the distance between the RIS and eavesdropper $j$. 

\noindent To calculate the ergodic secrecy rate, the ergodic up-link
rate for user $k$ and ergodic rate at the eavesdropper $j$ should
be derived, which will be considered in the following sub-sections.

\subsection{Ergodic Up-link rate of user $k$}

To calculate the ergodic user rate, maximum ratio combining (MRC)
is adopted at the BS. The beamforming matrix is given by $\mathbf{W}=\left(\mathbf{G}\Theta\mathbf{H}\right)^{H}$,
and thus $\mathbf{w}_{k}=\mathbf{h}_{r,k}^{H}\Theta^{H}\mathbf{G}^{H}$.
The signal to interference plus noise ratio (SINR) at the BS to decode
user $k$ signal can be written as 

\begin{equation}
\gamma_{b_{k}}=\frac{p_{k}\, L_{u_{k},b}\left|\mathbf{h}_{r,k}^{H}\Theta^{H}\mathbf{G}^{H}\mathbf{G}\Theta\bar{\Theta}\mathbf{h}_{r,k}\right|^{2}}{\stackrel[\underset{i\neq k}{i=1}]{K}{\sum}p_{i}\, L_{u_{i},b}\left|\mathbf{h}_{r,k}^{H}\Theta^{H}\mathbf{G}^{H}\mathbf{G}\Theta\bar{\Theta}\mathbf{h}_{r,i}\right|^{2}+\left\Vert \mathbf{h}_{r,k}^{H}\Theta^{H}\mathbf{G}^{H}\right\Vert ^{2}\sigma_{b}^{2}}.
\end{equation}

\begin{lem}
The ergodic up-link rate of user $k$ in passive RIS-aided MU-MISO
systems under Rician fading channels and with phase shift error can
be calculated by

\begin{equation}
\mathscr{E}\left\{ R_{b_{k}}\right\} \approx\log_{2}\left(1+\frac{p_{k}\, L_{u_{k},b}\xi_{k}}{\stackrel[\underset{i\neq k}{i=1}]{K}{\sum}p_{i}\, L_{u_{i},b}\varsigma_{i}+\upsilon_{k}\sigma_{b}^{2}}\right)
\end{equation}

\noindent where 

\[
\xi_{k}=\mathscr{E}\left\{ \left|\mathbf{h}_{r,k}^{H}\Theta^{H}\mathbf{G}^{H}\mathbf{G}\Theta\bar{\Theta}\mathbf{h}_{r,k}\right|^{2}\right\} =\frac{1}{\left(\rho_{b}+1\right)^{2}\left(\rho_{k}+1\right)^{2}}\left(a_{1}N^{2}+a_{2}NM^{2}+a_{3}NM+a_{4}N\right)
\]

$a_{1}=\left(\rho\left(\kappa\right)^{2}\left(\rho_{k}+\rho_{b}+1\right)^{2}+\left(1-\rho\left(\kappa\right)^{2}\right)\rho_{k}\rho_{b}^{2}+\rho_{b}^{2}\right)M^{2}$

$+\left(\left(\left(2\rho_{k}+3\rho_{b}+2-\rho_{k}\rho_{b}\right)\rho\left(\kappa\right)^{2}+\left(1+\rho_{k}\right)\rho_{b}\right)\rho_{b}\rho_{k}\left|f_{k}\right|^{2}+\left(\rho_{k}+\rho_{b}+2\right)^{2}\right.$

$-\left.\rho\left(\kappa\right)^{2}\left(\rho_{k}+\rho_{b}+1\right)^{2}-2\rho\left(\kappa\right)^{2}\rho_{k}\rho_{b}-2\right)M$

$+\rho\left(\kappa\right)^{2}\rho_{b}^{2}\rho_{k}^{2}\left|f_{k}\right|^{4}+2\left(\left(1-\rho\left(\kappa\right)^{2}\right)\left(\rho_{k}+\rho_{b}\right)+2\right)\rho_{b}\rho_{k}\left|f_{k}\right|^{2}$,

$a_{2}=\left(-\rho\left(\kappa\right)^{2}\rho_{k}\rho_{b}\left(1+\rho_{k}\right)M^{2}\right)+\left(\rho_{k}+\rho_{b}+1\right)\rho_{b}\rho_{r}+\left(\rho_{k}+\rho_{b}+1\right)^{2}-\left(\rho_{k}+1\right)\rho_{b}^{2}$,

$a_{3}=\left(\left(\left(\rho_{k}+1\right)\rho\left(\kappa\right)^{2}\right)+\left(\rho_{k}+1\right)\right)\rho_{b}\rho_{k}\left|f_{k}\right|^{2}-2\rho_{b}\rho_{k}\rho\left(\kappa\right)^{2}+2\rho_{b}\rho_{k}+2\rho_{k}+2\rho_{b}-1$,

$a_{4}=2\rho_{b}\rho_{r}\left|f_{k}\right|^{2}\left(1+\rho\left(\kappa\right)^{2}\right)$

and

\[
\varsigma_{i}=\mathscr{E}\left\{ \left|\mathbf{h}_{r,k}^{H}\Theta^{H}\mathbf{G}^{H}\mathbf{G}\Theta\bar{\Theta}\mathbf{h}_{r,i}\right|^{2}\right\} =\frac{1}{\left(\rho_{b}+1\right)^{2}\left(\rho_{k}+1\right)\left(\rho_{i}+1\right)}\left(b_{1}N^{2}+b_{2}NM^{2}+b_{3}NM\right)
\]

$b_{1}=\left(\rho_{i}+1-\rho\left(\kappa\right)^{2}\rho_{i}\right)M^{2}\rho_{b}^{2}$ 

$+M\left(\left(\rho_{i}+1-\rho\left(\kappa\right)^{2}\rho_{i}\right)\rho_{b}^{2}\rho_{k}\left|f_{k}\right|^{2}+\rho\left(\kappa\right)^{2}\rho_{b}^{2}\rho_{i}\left|f_{i}\right|^{2}+\left(\rho_{k}+2\rho_{b}+1\right)\left(\rho_{i}+1-\rho\left(\kappa\right)^{2}\rho_{i}\right)+\rho\left(\kappa\right)^{2}\rho_{i}\right)$

$+\left(2\rho_{b}\left|f_{i}\right|^{2}+\rho_{k}\left|\mathbf{\bar{h}}_{k}^{H}\mathbf{\bar{h}}_{i}\right|^{2}+2\rho_{b}\rho_{k}\textrm{Re}\left(f_{k}^{*}f_{i}\mathbf{\bar{h}}_{i}^{H}\mathbf{\bar{h}}_{k}\right)\right)\rho\left(\kappa\right)^{2}\rho_{i}$ 

$+\left(\rho\left(\kappa\right)^{2}\rho_{b}\rho_{i}\left|f_{i}\right|^{2}+2\rho_{i}\left(1-\rho\left(\kappa\right)^{2}\right)+2\right)\rho_{b}\rho_{k}\left|f_{k}\right|^{2}$

$b_{2}=\left(\left(\rho_{b}+1\right)\rho_{k}+\left(\rho_{b}+1\right)^{2}-\rho_{b}^{2}\right)\left(\rho_{i}+1\right)-\left(\rho_{b}+1\right)\rho_{b}\rho_{i}\rho\left(\kappa\right)^{2}-1$

$b_{3}=\left(\rho_{i}+1\right)\rho_{b}\rho_{k}\left|f_{k}\right|^{2}+\left(\rho_{k}+1\right)\rho\left(\kappa\right)^{2}\rho_{b}\rho_{i}\left|f_{i}\right|^{2}$

and

\[
\upsilon_{k}=\mathscr{E}\left\{ \left\Vert \mathbf{h}_{r,k}^{H}\Theta^{H}\mathbf{G}^{H}\right\Vert ^{2}\right\} =\frac{L_{u_{k},b}}{\left(\rho_{b}+1\right)\left(\rho_{k}+1\right)}\left(\rho_{b}\rho_{k}\left|f_{k}\right|^{2}+\left(\rho_{b}+\rho_{k}+1\right)M\right)
\]
\end{lem}
\begin{IEEEproof}
The proof is provided in Appendix A.
\end{IEEEproof}

\subsection{Ergodic Rate at Eavesdropper $j$}

The SINR at eavesdropper $j$ to decode user $k$ signal can be expressed
as

\begin{equation}
\gamma_{e_{j,k}}=\frac{p_{k}\,\left|d_{u_{k},r}^{-\frac{\alpha_{r}}{2}}d_{e_{j},r}^{-\frac{\alpha_{e}}{2}}\mathbf{h}_{e_{j},r}\Theta\bar{\Theta}\mathbf{h}_{r,k}+\, d_{e_{j},k}^{-\frac{\alpha_{e}}{2}}h_{e_{j},k}\right|^{2}}{\stackrel[\underset{i\neq k}{i=1}]{K}{\sum}p_{i}\,\left|d_{u_{i},r}^{-\frac{\alpha_{r}}{2}}d_{e_{j},r}^{-\frac{\alpha_{e}}{2}}\mathbf{h}_{e_{j},r}\Theta\bar{\Theta}\mathbf{h}_{r,i}+d_{e_{j},i}^{-\frac{\alpha_{e}}{2}}h_{e_{j},i}\right|^{2}+\sigma_{e_{j}}^{2}}.
\end{equation}

\begin{lem}
The ergodic rate at eavesdropper $j$ in up-link passive RIS-aided
MU-MISO systems under Rician fading channels and with phase shift
error can be calculated by

\begin{equation}
\mathscr{E}\left\{ R_{e_{j,k}}\right\} =\log_{2}\left(1+\frac{p_{k}\, x_{k}}{\stackrel[\underset{i\neq k}{i=1}]{K}{\sum}p_{i}\, y_{i}+\sigma_{e_{j}}^{2}}\right)
\end{equation}

\noindent where

$x_{k}=\left(d_{u_{k},r}^{-\alpha_{r}}d_{e_{j},r}^{-\alpha_{e}}\left(\frac{\rho_{e_{j}}}{\rho_{e_{j}}+1}\frac{\rho_{k}}{\rho_{k}+1}\left(M+\rho\left(\kappa\right)^{2}\xi\right)+\frac{\rho_{e_{j}}}{\rho_{e_{j}}+1}\frac{1}{\rho_{k}+1}M+\frac{\rho_{k}}{\rho_{k}+1}\frac{1}{\rho_{e_{j}}+1}M+\frac{1}{\rho_{e_{j}}+1}\frac{1}{\rho_{k}+1}M\right)+d_{e_{j},r}^{-\alpha_{e}}\right)$,
and 

$y_{i}=\, d_{u_{i},r}^{-\alpha_{r}}d_{e_{j},r}^{-\alpha_{e}}\left(\frac{\rho_{e_{j}}}{\rho_{e_{j}}+1}\frac{\rho_{i}}{\rho_{i}+1}\left(M+\rho\left(\kappa\right)^{2}\xi\right)+\frac{\rho_{e_{j}}}{\rho_{e_{j}}+1}\frac{1}{\rho_{i}+1}M+\frac{\rho_{i}}{\rho_{i}+1}\frac{1}{\rho_{e_{j}}+1}M+\frac{1}{\rho_{e_{j}}+1}\frac{1}{\rho_{i}+1}M+d_{e_{j},i}^{-\alpha_{e}}\right).$\end{lem}
\begin{IEEEproof}
The proof is provided in Appendix B.
\end{IEEEproof}
Finally, the ergodic secrecy rate in passive RIS scheme is presented
in the next theorem.

\textbf{\emph{Theorem 1}}\emph{. The ergodic secrecy rate in passive
RIS-aided MU-MISO systems under Rician fading channels and with phase
shift error can be calculated by}

\emph{
\begin{equation}
\hat{R}_{s}=\left[\log_{2}\left(1+\frac{p_{k}\, L_{u_{k},b}\xi_{k}}{\stackrel[\underset{i\neq k}{i=1}]{K}{\sum}p_{i}\, L_{u_{i},b}\varsigma_{i}+\upsilon_{k}\sigma_{b}^{2}}\right)-\log_{2}\left(1+\frac{p_{k}\, x_{k}}{\stackrel[\underset{i\neq k}{i=1}]{K}{\sum}p_{i}\, y_{i}+\sigma_{e_{j}}^{2}}\right)\right].^{+}
\end{equation}
}

\section{Active RIS \label{sec:Active-RIS}}

As we mentioned earlier, active RIS can adjust the phase shifts and
also amplify the reflected signal to compensate the attenuation from
the first link with extra power consumption. The signal reflected
by the active IRS can be written as 

\begin{equation}
\mathbf{y}_{r}=\tilde{\Theta}\stackrel[k=1]{K}{\sum}\sqrt{p_{k}\, d_{u_{k},r}^{-\alpha_{r}}}\mathbf{h}_{r,i}x_{i}+\tilde{\Theta}\mathbf{n}_{r}
\end{equation}

\noindent where $\mathbf{n}_{r}$ is the noise at RIS elements $\mathbf{n}_{r}\sim CN\left(0,\sigma_{r}^{2}\mathbf{I}\right)$.
In this case $\tilde{\Theta}=\bar{\Theta}\Theta$ where $\Theta=\textrm{diag}\left(\theta\right)$,
and $\theta=\left[\theta_{1},......,\theta_{M}\right]^{T}$ with $\theta_{m}=\varrho_{m}e^{j\varphi_{m}}$,
$\varrho_{m}>1$ and $\varphi_{m}\text{\ensuremath{\in}}[0,2\pi)$
represents the amplification factor and phase shift coefficient, respectively,
at element $m$. For simplicity, we assume that $\varrho_{m}=\varrho$
and then define $\Theta=\varrho\textrm{diag}\left\{ e^{j\varphi_{1}},....,e^{j\varphi_{M}}\right\} $.
The active RIS amplification power can be expressed as 

\begin{equation}
P_{r}=\left(\stackrel[k=1]{K}{\sum}\frac{p_{k}}{d_{u_{k},r}^{\alpha_{r}}}\mathscr{E}\left\{ \left\Vert \tilde{\Theta}\mathbf{h}_{r,i}\right\Vert ^{2}\right\} +\mathscr{E}\left\{ \left\Vert \tilde{\Theta}\mathbf{n}_{r}\right\Vert ^{2}\right\} \right)=\left(\stackrel[k=1]{K}{\sum}\frac{p_{k}}{d_{u_{k},r}^{\alpha_{r}}}M\varrho^{2}+M\varrho^{2}\sigma_{r}^{2}\right)
\end{equation}

\noindent where $\mathscr{E}\left\Vert \tilde{\Theta}\mathbf{n}_{r}\right\Vert ^{2}=M\varrho^{2}\sigma_{r}^{2}$,
and $\mathscr{E}\left\Vert \tilde{\Theta}\mathbf{h}_{r,i}\right\Vert ^{2}=\frac{\varrho^{2}}{\rho_{i}+1}\left(\rho_{i}\mathscr{E}\left(\mathbf{\bar{h}}_{r,i}^{H}\mathbf{\bar{h}}_{r,i}\right)+\mathscr{E}\left(\mathbf{\tilde{h}}_{r,i}^{H}\mathbf{\tilde{h}}_{r,i}\right)\right)=\frac{\varrho^{2}}{\rho_{i}+1}\left(\rho_{i}M+M\right)=M\varrho^{2}$.
Thus, the amplification factor for each element on the active RIS
is given by 

\begin{equation}
\varrho=\sqrt{\frac{P_{r}}{M\left(\stackrel[k=1]{K}{\sum}\frac{p_{k}}{d_{u_{k},r}^{\alpha_{r}}}+\sigma_{r}^{2}\right)}}.\label{eq:24}
\end{equation}

\noindent By applying the receive beamforming vector $\mathbf{w}_{k}$
at the BS, the received signal of user $k$ is 

\begin{equation}
y_{b,k}=\sqrt{p_{k}\, L_{u_{k},b}}\mathbf{w}_{k}\mathbf{G}\bar{\Theta}\Theta\mathbf{h}_{r,k}x_{k}+\stackrel[\underset{i\neq k}{i=1}]{K}{\sum}\sqrt{p_{i}\, L_{u_{i},b}}\mathbf{w}_{k}\mathbf{G}\bar{\Theta}\Theta\mathbf{h}_{r,i}x_{i}+\sqrt{d_{r,b}^{-\alpha_{r}}}\mathbf{w}_{k}\mathbf{G}\bar{\Theta}\Theta\mathbf{n}_{r}+\mathbf{w}_{k}\mathbf{n}_{b}.
\end{equation}

On the other hand, the received signal at eavesdropper $j$ to detect
user $k$ signal is

\[
y_{e_{j,k}}=\sqrt{p_{k}}x_{k}\left(\sqrt{d_{e_{j},k}^{-\alpha_{e}}}h_{e_{j},k}+\sqrt{L_{u_{k},e_{j}}}\mathbf{h}_{e_{j},r}\bar{\Theta}\Theta\mathbf{h}_{r,k}\right)
\]

\begin{equation}
+\stackrel[\underset{i\neq k}{i=1}]{K}{\sum}\sqrt{p_{i}}x_{i}\left(\sqrt{\, d_{e_{j},i}^{-\alpha_{e}}}h_{e_{j},i}+\stackrel[\underset{i\neq k}{i=1}]{K}{\sum}\sqrt{L_{u_{i},e_{j}}}\mathbf{h}_{e_{j},r}\bar{\Theta}\Theta\mathbf{h}_{r,i}\right)+\sqrt{d_{e_{j},r}^{-\alpha_{e}}}\mathbf{h}_{e_{j},r}\bar{\Theta}\Theta\mathbf{n}_{r}+n_{e_{j}}.
\end{equation}

\subsection{Ergodic Up-link rate of user $k$}

Applying MRC beamforming at the BS, the SINRs at the BS to decode
user $k$ signal can be expressed as

\begin{equation}
\gamma_{b_{k}}=\frac{p_{k}\, L_{u_{k},b}\left|\mathbf{h}_{r,k}^{H}\Theta^{H}\mathbf{G}^{H}\mathbf{G}\Theta\bar{\Theta}\mathbf{h}_{r,k}\right|^{2}}{\stackrel[\underset{i\neq k}{i=1}]{K}{\sum}p_{i}\, L_{u_{i},b}\left|\mathbf{h}_{r,k}^{H}\Theta^{H}\mathbf{G}^{H}\mathbf{G}\Theta\bar{\Theta}\mathbf{h}_{r,i}\right|^{2}+d_{r,b}^{-\alpha_{r}}\left\Vert \mathbf{h}_{r,k}^{H}\Theta^{H}\mathbf{G}^{H}\mathbf{G}\bar{\Theta}\Theta\right\Vert ^{2}\sigma_{r}^{2}+\left\Vert \mathbf{h}_{r,k}^{H}\Theta^{H}\mathbf{G}^{H}\right\Vert ^{2}\sigma_{b}^{2}}.
\end{equation}

\begin{lem}
The ergodic up-link rate of user $k$ in active RIS-aided MU-MISO
systems under Rician fading channels and with phase shift error can
be calculated by

\begin{equation}
\mathscr{E}\left\{ R_{b_{k}}\right\} \approx\log_{2}\left(1+\frac{p_{k}\, L_{u_{k},b}\xi_{k}\varrho^{4}}{\stackrel[\underset{i\neq k}{i=1}]{K}{\sum}p_{i}\, L_{u_{i},b}\varsigma_{i}\varrho^{4}+\varrho^{4}d_{r,b}^{-\alpha_{r}}\sigma_{r}^{2}\nu_{k}+\varrho^{2}\upsilon_{k}\sigma_{b}^{2}}\right)\label{eq:33}
\end{equation}

\noindent where

\begin{equation}
\nu_{k}=\mathscr{E}\left\Vert \mathbf{h}_{r,k}^{H}\Theta^{H}\mathbf{G}^{H}\mathbf{G}\bar{\Theta}\Theta\right\Vert ^{2}=\frac{1}{\left(\rho_{b}+1\right)\sqrt{\left(\rho_{k}+1\right)}}\left(X_{1}+X_{2}\right)
\end{equation}

and $X_{1}=\mathscr{E}\left\{ \left|\Delta_{1,1}\right|^{2}\right\} +\mathscr{E}\left\{ \left|\Delta_{1,2}\right|^{2}\right\} +\mathscr{E}\left\{ \left|\Delta_{1,3}\right|^{2}\right\} +\mathscr{E}\left\{ \left|\Delta_{1,4}\right|^{2}\right\} +\mathscr{E}\left\{ \Delta_{1,1}\Delta_{1,4}^{*}\right\} $

\[
\mathscr{E}\left\{ \left|\Delta_{1,1}\right|^{2}\right\} =\rho_{b}^{2}\rho_{k}\left|\left(a_{M}^{H}\left(\phi_{kr}^{a},\phi_{kr}^{e}\right)\Theta^{H}a_{M}^{H}\left(\phi_{r}^{a},\phi_{r}^{e}\right)a_{N}^{H}\left(\phi_{b}^{a},\phi_{b}^{e}\right)a_{N}\left(\phi_{b}^{a},\phi_{b}^{e}\right)\right)\right|^{2}\times M,
\]

\[
\mathscr{E}\left\{ \left|\Delta_{1,2}\right|^{2}\right\} =\rho_{b}\rho_{k}\left|a_{M}^{H}\left(\phi_{kr}^{a},\phi_{kr}^{e}\right)\Theta^{H}a_{M}\left(\phi_{r}^{a},\phi_{r}^{e}\right)\right|^{2}NM,
\]

\[
\mathscr{E}\left\{ \left|\Delta_{1,3}\right|^{2}\right\} =\rho_{b}\rho_{k}MN\left(\rho\left(\kappa\right)^{2}M+\left(1-\rho\left(\kappa\right)^{2}\right)M\right),\:\mathscr{E}\left\{ \left|\Delta_{1,4}\right|^{2}\right\} =\rho_{k}\left(N^{2}M+NM^{2}\right),
\]

\[
\mathscr{E}\left\{ \Delta_{1,1}\Delta_{1,4}^{*}\right\} =\rho_{b}\rho_{k}\left(a_{M}^{H}\left(\phi_{kr}^{a},\phi_{kr}^{e}\right)\Theta^{H}a_{M}^{H}\left(\phi_{r}^{a},\phi_{r}^{e}\right)a_{N}^{H}\left(\phi_{b}^{a},\phi_{b}^{e}\right)a_{N}\left(\phi_{b}^{a},\phi_{b}^{e}\right)\right)
\]

\[
\times\left(a_{M}\left(\phi_{r}^{a},\phi_{r}^{e}\right)\Theta\right)\rho_{k}a_{M}^{H}\left(\phi_{kr}^{a},\phi_{kr}^{e}\right)\Theta^{H}N\Theta,
\]

and $X_{2}=\mathscr{E}\left\{ \left|\Delta_{2,1}\right|^{2}\right\} +\mathscr{E}\left\{ \left|\Delta_{2,2}\right|^{2}\right\} +\mathscr{E}\left\{ \left|\Delta_{2,3}\right|^{2}\right\} +\mathscr{E}\left\{ \left|\Delta_{2,4}\right|^{2}\right\} +\mathscr{E}\left\{ \Delta_{2,1}\Delta_{2,4}^{*}\right\} $

\[
\mathscr{E}\left\{ \left|\Delta_{2,1}\right|^{2}\right\} =\rho_{b}^{2}\left\Vert \Theta^{H}a_{M}^{H}\left(\phi_{r}^{a},\phi_{r}^{e}\right)a_{N}^{H}\left(\phi_{b}^{a},\phi_{b}^{e}\right)a_{N}\left(\phi_{b}^{a},\phi_{b}^{e}\right)a_{M}\left(\phi_{r}^{a},\phi_{r}^{e}\right)\Theta\right\Vert _{F}^{2},
\]

\[
\mathscr{E}\left\{ \left|\Delta_{2,2}\right|^{2}\right\} =\rho_{b}\left\Vert \Theta^{H}a_{M}\left(\phi_{r}^{a},\phi_{r}^{e}\right)\right\Vert ^{2}NM,
\]

\[
\mathscr{E}\left\{ \left|\Delta_{2,3}\right|^{2}\right\} =\rho_{b}MN\left(\left\Vert a_{M}^{H}\left(\phi_{r}^{a},\phi_{r}^{e}\right)\Theta\bar{\Theta}\right\Vert ^{2}\right)=\rho_{b}M^{2}N,\quad\mathscr{E}\left\{ \left|\Delta_{2,4}\right|^{2}\right\} =\rho_{k}^{2}\left(N^{2}M+NM^{2}\right),
\]

\[
\mathscr{E}\left\{ \Delta_{2,1}\Delta_{2,4}^{*}\right\} =\rho_{k}\left(\Theta^{H}a_{M}^{H}\left(\phi_{r}^{a},\phi_{r}^{e}\right)a_{N}^{H}\left(\phi_{b}^{a},\phi_{b}^{e}\right)a_{N}\left(\phi_{b}^{a},\phi_{b}^{e}\right)\right)\left(a_{M}\left(\phi_{r}^{a},\phi_{r}^{e}\right)\Theta\right)\rho_{k}\Theta N\Theta^{H}.
\]
\end{lem}
\begin{IEEEproof}
The proof is provided in Appendix C.
\end{IEEEproof}

\subsection{Ergodic Rate at Eavesdropper $j$}

The SINR at eavesdropper $j$ to decode user $k$ signal in this scenario
can be written as 

\begin{equation}
\gamma_{e_{j,k}}=\frac{p_{k}\,\left|d_{u_{k},r}^{-\frac{\alpha_{r}}{2}}d_{e_{j},r}^{-\frac{\alpha_{e}}{2}}\mathbf{h}_{e_{j},r}\Theta\bar{\Theta}\mathbf{h}_{r,k}+\, d_{e_{j},k}^{-\frac{\alpha_{e}}{2}}h_{e_{j},k}\right|^{2}}{\stackrel[\underset{i\neq k}{i=1}]{K}{\sum}p_{i}\,\left|d_{u_{i},r}^{-\frac{\alpha_{r}}{2}}d_{e_{j},r}^{-\frac{\alpha_{e}}{2}}\mathbf{h}_{e_{j},r}\Theta\bar{\Theta}\mathbf{h}_{r,i}+d_{e_{j},i}^{-\frac{\alpha_{e}}{2}}h_{e_{j},i}\right|^{2}+d_{e_{j},r}^{-\alpha_{e}}\left\Vert \mathbf{h}_{e_{j},r}\bar{\Theta}\Theta\right\Vert ^{2}\sigma_{r}^{2}+\sigma_{e_{j}}^{2}}.
\end{equation}

\begin{lem}
The ergodic rate at eavesdropper $j$ in up-link active RIS-aided
MU-MISO systems under Rician fading channels and with phase shift
error can be calculated by

\begin{equation}
\mathscr{E}\left\{ R_{e_{j,k}}\right\} =\log_{2}\left(1+\frac{p_{k}\, x_{j}}{\stackrel[\underset{i\neq k}{i=1}]{K}{\sum}p_{i}y_{i}+z_{j}\sigma_{r}^{2}+\sigma_{e_{j}}^{2}}\right)
\end{equation}

\noindent where

$x_{j}=\left(d_{u_{k},r}^{-\alpha_{r}}d_{e_{j},r}^{-\alpha_{e}}\varrho^{2}\left(\frac{\rho_{e_{j}}}{\rho_{e_{j}}+1}\frac{\rho_{k}}{\rho_{k}+1}\left(M+\rho\left(\kappa\right)^{2}\xi\right)+\frac{\rho_{e_{j}}}{\rho_{e_{j}}+1}\frac{1}{\rho_{k}+1}M+\frac{\rho_{k}}{\rho_{k}+1}\frac{1}{\rho_{e_{j}}+1}M+\frac{1}{\rho_{e_{j}}+1}\frac{1}{\rho_{k}+1}M\right)+d_{e_{j},r}^{-\alpha_{e}}\right)$, 

$y_{k}=\, d_{u_{i},r}^{-\alpha_{r}}d_{e_{j},r}^{-\alpha_{e}}\varrho^{2}\left(\frac{\rho_{e_{j}}}{\rho_{e_{j}}+1}\frac{\rho_{i}}{\rho_{i}+1}\left(M+\rho\left(\kappa\right)^{2}\xi\right)+\frac{\rho_{e_{j}}}{\rho_{e_{j}}+1}\frac{1}{\rho_{i}+1}M+\frac{\rho_{i}}{\rho_{i}+1}\frac{1}{\rho_{e_{j}}+1}M+\frac{1}{\rho_{e_{j}}+1}\frac{1}{\rho_{i}+1}M+d_{e_{j},i}^{-\alpha_{e}}\right)$

which have been derived in Appemdix B, and

$z_{j}=d_{e_{j},r}^{-\alpha_{e}}\mathscr{E}\left\Vert \mathbf{h}_{e_{j},r}\tilde{\Theta}\right\Vert ^{2}=d_{e_{j},r}^{-\alpha_{e}}\frac{\varrho^{2}}{\rho_{e_{j,}r}+1}\left(\rho_{e_{j,}r}\mathscr{E}\left(\mathbf{\bar{h}}_{e_{j},r}^{H}\mathbf{\bar{h}}_{e_{j},r}\right)+\mathscr{E}\left(\mathbf{\tilde{h}}_{e_{j},r}^{H}\mathbf{\tilde{h}}_{e_{j},r}\right)\right)$

$=d_{e_{j},r}^{-\alpha_{e}}\frac{\varrho^{2}}{\rho_{e_{j,}r}+1}\left(\rho_{e_{j,}r}M+M\right)=d_{e_{j},r}^{-\alpha_{e}}M\varrho^{2}.$
\end{lem}
The ergodic secrecy rate in active RIS scheme is presented in the
following Theorem. 

\textbf{\emph{Theorem }}\emph{2. The ergodic secrecy rate in active
RIS-aided MU-MISO systems under Rician fading channels and with phase
shift error can be calculated by}

\emph{
\begin{equation}
\hat{R}_{s}=\left[\log_{2}\left(1+\frac{p_{k}\, L_{u_{k},b}\xi_{k}\varrho^{4}}{\stackrel[\underset{i\neq k}{i=1}]{K}{\sum}p_{i}\, L_{u_{i},b}\varsigma_{i}\varrho^{4}+\varrho^{4}d_{r,b}^{-\alpha_{r}}\sigma_{r}^{2}\nu_{k}+\varrho^{2}\upsilon_{k}\sigma_{b}^{2}}\right)-\log_{2}\left(1+\frac{p_{k}\, x_{j}}{\stackrel[\underset{i\neq k}{i=1}]{K}{\sum}p_{i}y_{i}+z_{j}\sigma_{r}^{2}+\sigma_{e_{j}}^{2}}\right)\right]^{+}.
\end{equation}
}

\section{EH RIS\label{sec:EH-RIS}}

Following the recent works in \cite{Ref19,Ref20,Ref21,Ref22,Ref23},
in this section, the RIS is an energy constrained node and it can
harvest RF energy to support its operation. Thus, in this scenario
the whole operation time block, $T$, is split into two time periods,
the energy transfer (ET) slot and the information transfer (IT) slot.
During the ET slot, the BS transmits energy signals to the RIS to
support its operation. During the IT slot, the users deliver their
messages to the BS through the RIS. We denote $\tau T$ as the time
duration for the ET, and $\left(1-\tau\right)T$ as the time duration
for IT. The received signals at the RIS in the first sub-slot is expressed
as

\begin{equation}
y_{r}=\sqrt{P_{b}}\mathbf{G}_{p}\mathbf{W}_{p}\mathbf{x}_{p}+\mathbf{n}_{r}
\end{equation}

\noindent where $P_{b}$ is the BS power, $\mathbf{G}_{p}=\left(\sqrt{\frac{\rho_{p}}{\rho_{p}+1}}\mathbf{\bar{G}}_{p}+\sqrt{\frac{1}{\rho_{p}+1}}\mathbf{\tilde{G}}_{p}\right)$
is the BS-RIS channel in the ET slot, $\mathbf{W}_{p}$ is the precoding
matrix and $\mathbf{x}_{p}$ is the energy signals vector. Using the
maximum ratio transmission (MRT) scheme, the harvested power at the
RIS can be expressed as $P_{r}=\frac{\eta_{eff}\tau P_{b}\left\Vert \mathbf{G}_{p}\right\Vert _{F}^{2}}{1-\tau}$,
which can be written as $P_{r}=\frac{\eta_{eff}\tau P_{b}\textrm{Tr}\left(G_{b}G_{b}^{H}\right)}{1-\tau}$
where $\eta_{eff}$ is the efficiency of EH. Since $G_{b}G_{b}^{H}$
has Wishart distribution, the average harvested power can be written
as

\begin{equation}
P_{r}=\frac{\eta_{eff}\tau P_{b}\mathscr{E}\left\{ \textrm{Tr}\left(G_{b}G_{b}^{H}\right)\right\} }{1-\tau}=\frac{\eta_{eff}\tau P_{b}NM}{1-\tau}.\label{eq:38}
\end{equation}

\noindent By substituting (\ref{eq:38}) into (\ref{eq:24}), the
amplification factor for each element on the RIS in this case is given
by 

\begin{equation}
\hat{\varrho}=\sqrt{\frac{\eta_{eff}\tau P_{b}NM}{M\left(1-\tau\right)\left(\stackrel[k=1]{K}{\sum}\frac{p_{k}}{d_{u_{k},r}^{\alpha_{r}}}+\sigma_{r}^{2}\right)}}.
\end{equation}

\subsection{Ergodic Up-link rate of user $k$}

Applying MRC beamforming at the BS, the SINR at the BS to decode
user $k$ signal can be expressed as

\begin{equation}
\gamma_{b_{k}}=\frac{p_{k}\, L_{u_{k},b}\left|\mathbf{h}_{r,k}^{H}\Theta^{H}\mathbf{G}^{H}\mathbf{G}\Theta\bar{\Theta}\mathbf{h}_{r,k}\right|^{2}}{\stackrel[\underset{i\neq k}{i=1}]{K}{\sum}p_{i}\, L_{u_{i},b}\left|\mathbf{h}_{r,k}^{H}\Theta^{H}\mathbf{G}^{H}\mathbf{G}\Theta\bar{\Theta}\mathbf{h}_{r,i}\right|^{2}+d_{r,b}^{-\alpha_{r}}\left\Vert \mathbf{h}_{r,k}^{H}\Theta^{H}\mathbf{G}^{H}\mathbf{G}\bar{\Theta}\Theta\right\Vert ^{2}\sigma_{r}^{2}+\left\Vert \mathbf{h}_{r,k}^{H}\Theta^{H}\mathbf{G}^{H}\right\Vert ^{2}\sigma_{b}^{2}}.
\end{equation}

\begin{lem}
The ergodic up-link rate of user $k$ in EH RIS-aided MU-MISO systems
under Rician fading channels and with phase shift error can be calculated
by

\begin{equation}
\mathscr{E}\left\{ R_{b_{k}}\right\} \approx\left(1-\tau\right)\log_{2}\left(1+\frac{p_{k}\, L_{u_{k},b}\xi_{k}\hat{\varrho}^{4}}{\stackrel[\underset{i\neq k}{i=1}]{K}{\sum}p_{i}\, L_{u_{i},b}\varsigma_{i}\hat{\varrho}^{4}+\hat{\varrho}^{4}d_{r,b}^{-\alpha_{r}}\sigma_{r}^{2}\nu_{k}+\hat{\varrho}^{2}\upsilon_{k}\sigma_{b}^{2}}\right).\label{eq:33-1}
\end{equation}
\end{lem}
\begin{IEEEproof}
This expression can be obtained by following same derivation in Appendix
C.
\end{IEEEproof}

\subsection{Ergodic Rate at Eavesdropper $j$}

The SINR at eavesdropper $j$ to decode user $k$ signal is given
by 

\begin{equation}
\gamma_{e_{j,k}}=\frac{p_{k}\,\left|d_{u_{k},r}^{-\frac{\alpha_{r}}{2}}d_{e_{j},r}^{-\frac{\alpha_{e}}{2}}\mathbf{h}_{e_{j},r}\Theta\bar{\Theta}\mathbf{h}_{r,k}+\, d_{e_{j},k}^{-\frac{\alpha_{e}}{2}}h_{e_{j},k}\right|^{2}}{\stackrel[\underset{i\neq k}{i=1}]{K}{\sum}p_{i}\,\left|d_{u_{i},r}^{-\frac{\alpha_{r}}{2}}d_{e_{j},r}^{-\frac{\alpha_{e}}{2}}\mathbf{h}_{e_{j},r}\Theta\bar{\Theta}\mathbf{h}_{r,i}+d_{e_{j},i}^{-\frac{\alpha_{e}}{2}}h_{e_{j},i}\right|^{2}+d_{e_{j},r}^{-\alpha_{e}}\left\Vert \mathbf{h}_{e_{j},r}\bar{\Theta}\Theta\right\Vert ^{2}\sigma_{r}^{2}+\sigma_{e_{j}}^{2}}.
\end{equation}

\begin{lem}
The ergodic rate at eavesdropper $j$ in up-link EH RIS-aided MU-MISO
systems under Rician fading channels and with phase shift error can
be calculated by

\begin{equation}
\mathscr{E}\left\{ R_{e_{j,k}}\right\} =\left(1-\tau\right)\log_{2}\left(1+\frac{p_{k}\,\hat{x}_{j}}{\stackrel[\underset{i\neq k}{i=1}]{K}{\sum}p_{i}\hat{y}_{i}+\hat{z}_{j}\sigma_{r}^{2}+\sigma_{e_{j}}^{2}}\right)
\end{equation}

\noindent where

$\hat{x}_{j}=\left(d_{u_{k},r}^{-\alpha_{r}}d_{e_{j},r}^{-\alpha_{e}}\hat{\varrho}^{2}\left(\frac{\rho_{e_{j}}}{\rho_{e_{j}}+1}\frac{\rho_{k}}{\rho_{k}+1}\left(M+\rho\left(\kappa\right)^{2}\xi\right)+\frac{\rho_{e_{j}}}{\rho_{e_{j}}+1}\frac{1}{\rho_{k}+1}M+\frac{\rho_{k}}{\rho_{k}+1}\frac{1}{\rho_{e_{j}}+1}M+\frac{1}{\rho_{e_{j}}+1}\frac{1}{\rho_{k}+1}M\right)+d_{e_{j},r}^{-\alpha_{e}}\right)$, 

$\hat{y}_{i}=\, d_{u_{i},r}^{-\alpha_{r}}d_{e_{j},r}^{-\alpha_{e}}\hat{\varrho}^{2}\left(\frac{\rho_{e_{j}}}{\rho_{e_{j}}+1}\frac{\rho_{i}}{\rho_{i}+1}\left(M+\rho\left(\kappa\right)^{2}\xi\right)+\frac{\rho_{e_{j}}}{\rho_{e_{j}}+1}\frac{1}{\rho_{i}+1}M+\frac{\rho_{i}}{\rho_{i}+1}\frac{1}{\rho_{e_{j}}+1}M+\frac{1}{\rho_{e_{j}}+1}\frac{1}{\rho_{i}+1}M+d_{e_{j},i}^{-\alpha_{e}}\right)$

$\hat{z}_{j}=d_{e_{j},r}^{-\alpha_{e}}M\hat{\varrho}^{2}$,

\noindent which have been derived in the previous section.
\end{lem}
Finally, the ergodic secrecy rate in EH RIS scheme is presented in
the next Theorem. 

\textbf{\emph{Theorem }}\emph{3. The ergodic secrecy rate of user
$k$ in EH active RIS-aided MU-MISO systems under Rician fading channels
and with phase shift error can be calculated by}

\emph{
\begin{eqnarray}
\hat{R}_{s} & = & \left[\left(1-\tau\right)\log_{2}\left(1+\frac{p_{k}\, L_{u_{k},b}\xi_{k}\hat{\varrho}^{4}}{\stackrel[\underset{i\neq k}{i=1}]{K}{\sum}p_{i}\, L_{u_{i},b}\varsigma_{i}\hat{\varrho}^{4}+\hat{\varrho}^{4}d_{r,b}^{-\alpha_{r}}\sigma_{r}^{2}\nu_{k}+\hat{\varrho}^{2}\upsilon_{k}\sigma_{b}^{2}}\right)\right.\nonumber \\
 &  & \left.-\left(1-\tau\right)\log_{2}\left(1+\frac{p_{k}\,\hat{x}_{j}}{\stackrel[\underset{i\neq k}{i=1}]{K}{\sum}p_{i}\hat{y}_{i}+\hat{z}_{j}\sigma_{r}^{2}+\sigma_{e_{j}}^{2}}\right)\right]^{+}.
\end{eqnarray}
}

\section{System Design\label{sec:System-Design}}

In this section, based on the derived analytical expressions, we first
design the phase shifts of the RIS configurations considered in this
work. Then, the best RIS configuration selection scheme is presented.

\subsection{Phase Shift Optimization }

The secrecy rate expressions presented in Theorems, 1, 2 and 3, show
that the secrecy performance relies on the phase shifts of the RIS
elements. In this work, it is assumed that the CSI of the eavesdroppers
is unknown at the BS/RIS (only channel distribution known). Therefore,
to enhance the system performance, the RIS phase shifts can be optimized
by maximizing the achievable ergodic sum rate. Since the phase shift
at each unit of the RIS lies in the range of $[0;2\pi)$, the phase
shift optimization problem can be formulated as

\begin{eqnarray}
\underset{\Theta}{\max} & \stackrel[i=1]{K}{\sum} & \hat{R}_{b_{i}}\nonumber \\
s.t & \theta_{m}\in\left[\left.0,2\pi\right),\right. & \lor m.\label{eq:18}
\end{eqnarray}

Due to the complicated formula of the ergodic sum rate, it is difficult
to optimize (\ref{eq:18}) based on the conventional techniques. However,
GA-based methods can be employed to solve this optimization problem.
Due to the page limitation, we refer readers to \cite{Ref5} for more
details about the GA methods.

As an efficient suboptimal solution, the RIS phase shifts can be aligned
to user $k$, who transmits the confidential message. This presents
a simple sub-optimal solution for enhancing the secrecy rate \cite{Ref5}.
Accordingly, the phase shifts should be

\begin{equation}
\theta_{m}=-2\pi\frac{d}{\lambda}\left(x_{m}t_{k}+y_{m}l_{k}\right),\, t_{k}=\sin\phi_{kr}^{a}\sin\phi_{kr}^{e}-\sin\phi_{t}^{a}\sin\phi_{t}^{e},\, l_{k}=\cos\phi_{kr}^{e}-\cos\phi_{t}^{e}.
\end{equation}

\noindent

\subsection{RIS Configuration Selection Scheme }

Based on the required secrecy rate $\left(r_{s}\right)$ and amount
of the power available at user $k$, and the RIS, we can decide which
system configuration, i.e., passive RIS, active RIS or EH RIS, should
be selected.

\textbf{A)} If user $k$ has sufficient amount of power to achieve
the target secrecy rate, in this case passive RIS can be implemented.
Based on the secrecy rate expression provided in Theorem 1, the required
user $k$ power, $p_{k}$, to achieve the target secrecy rate, $r_{s}$,
can be obtained by solving 

\begin{equation}
r_{s}=\log_{2}\left(1+\frac{p_{k}\, L_{u_{k},b}\xi_{k}}{\stackrel[\underset{i\neq k}{i=1}]{K}{\sum}p_{i}\, L_{u_{i},b}\varsigma_{i}+\upsilon_{k}\sigma_{b}^{2}}\right)-\log_{2}\left(1+\frac{p_{k}\, x_{k}}{\stackrel[\underset{i\neq k}{i=1}]{K}{\sum}p_{i}\, y_{i}+\sigma_{e_{j}}^{2}}\right)
\end{equation}

\noindent which can be found as

\begin{equation}
p_{k}=\frac{p_{1}-p_{2}}{p_{3}-p_{4}}\label{eq:49}
\end{equation}

\noindent where $p_{1}=\frac{\stackrel[\underset{i\neq k}{i=1}]{K}{\sum}p_{i}\, L_{u_{i},b}\varsigma_{i}+\upsilon_{k}\sigma_{b}^{2}}{\stackrel[\underset{i\neq k}{i=1}]{K}{\sum}p_{i}\, L_{u_{i},b}\varsigma_{i}+\upsilon_{k}\sigma_{b}^{2}}$,
$p_{2}=\frac{2^{r_{s}}\stackrel[\underset{i\neq k}{i=1}]{K}{\sum}p_{i}\, y_{i}+2^{r_{s}}\sigma_{e_{j}}^{2}}{\stackrel[\underset{i\neq k}{i=1}]{K}{\sum}p_{i}\, y_{i}+\sigma_{e_{j}}^{2}}$,
$p_{3}=\frac{2^{r_{s}}\, x_{k}}{\stackrel[\underset{i\neq k}{i=1}]{K}{\sum}p_{i}\, y_{i}+\sigma_{e_{j}}^{2}}$
and $p_{4}=\frac{\, L_{u_{k},b}\xi_{k}}{\stackrel[\underset{i\neq k}{i=1}]{K}{\sum}p_{i}\, L_{u_{i},b}\varsigma_{i}+\upsilon_{k}\sigma_{b}^{2}}$.

\textbf{B)} If user $k$ has limited amount of power, e.g., the user
power, $p_{k}$, is less than the power required in (\ref{eq:49}).
In this case active RIS can be implemented to provide the target secrecy
rate. Based on the secrecy rate expression provided in Theorem 2,
the required RIS power, $\varrho$ or $P_{r}$ ,to achieve the target
secrecy rate, $r_{s}$, can be obtained by solving 

\[
r_{s}=\log_{2}\left(1+\frac{p_{k}\, L_{u_{k},b}\xi_{k}\varrho^{2}}{\stackrel[\underset{i\neq k}{i=1}]{K}{\sum}p_{i}\, L_{u_{i},b}\varsigma_{i}\varrho^{2}+\varrho^{2}d_{r,b}^{-\alpha_{r}}\sigma_{r}^{2}\nu_{k}+\upsilon_{k}\sigma_{b}^{2}}\right)
\]

\begin{equation}
-\log_{2}\left(1+\frac{p_{k}\varrho^{2}x_{1}+p_{k}x_{2}}{\stackrel[\underset{i\neq k}{i=1}]{K}{\sum}p_{i}\varrho^{2}y_{1i}+\stackrel[\underset{i\neq k}{i=1}]{K}{\sum}p_{i}y_{2i}+z_{1}\varrho^{2}\sigma_{r}^{2}+\sigma_{e_{j}}^{2}}\right)
\end{equation}

\noindent where $x_{1}=d_{u_{k},r}^{-\alpha_{r}}d_{e_{j},r}^{-\alpha_{e}}\left(\frac{\rho_{e_{j}}}{\rho_{e_{j}}+1}\frac{\rho_{k}}{\rho_{k}+1}\left(M+\rho\left(\kappa\right)^{2}\xi\right)+\frac{\rho_{e_{j}}}{\rho_{e_{j}}+1}\frac{1}{\rho_{k}+1}M+\frac{\rho_{k}}{\rho_{k}+1}\frac{1}{\rho_{e_{j}}+1}M+\frac{1}{\rho_{e_{j}}+1}\frac{1}{\rho_{k}+1}M\right)$,
$x_{2}=d_{e_{j},r}^{-\alpha_{e}}$, 

$y_{1i}=\, d_{u_{i},r}^{-\alpha_{r}}d_{e_{j},r}^{-\alpha_{e}}\left(\frac{\rho_{e_{j}}}{\rho_{e_{j}}+1}\frac{\rho_{i}}{\rho_{i}+1}\left(M+\rho\left(\kappa\right)^{2}\xi\right)+\frac{\rho_{e_{j}}}{\rho_{e_{j}}+1}\frac{1}{\rho_{i}+1}M+\frac{\rho_{i}}{\rho_{i}+1}\frac{1}{\rho_{e_{j}}+1}M+\frac{1}{\rho_{e_{j}}+1}\frac{1}{\rho_{i}+1}M\right)$,
$y_{2i}=d_{e_{j},i}^{-\alpha_{e}}$,

and $z_{1}=d_{e_{j},r}^{-\alpha_{e}}M.$ After some simplifications,
the last equation can be expressed as 

\begin{equation}
\varrho^{4}\left(q_{1}-q_{3}\right)+\varrho^{2}\left(q_{2}-q_{4}-q_{5}+q_{7}\right)+\left(q_{8}-q_{6}\right)=0
\end{equation}

\noindent where

$q_{1}=\stackrel[\underset{i\neq k}{i=1}]{K}{\sum}p_{i}\, L_{u_{i},b}\varsigma_{i}\stackrel[\underset{i\neq k}{i=1}]{K}{\sum}p_{i}y_{1i}+d_{r,b}^{-\alpha_{r}}\sigma_{r}^{2}\nu_{k}\stackrel[\underset{i\neq k}{i=1}]{K}{\sum}p_{i}y_{1i}+\stackrel[\underset{i\neq k}{i=1}]{K}{\sum}p_{i}\, L_{u_{i},b}\varsigma_{i}z_{1}\sigma_{r}^{2}+d_{r,b}^{-\alpha_{r}}\sigma_{r}^{2}\nu_{k}z_{1}\sigma_{r}^{2}+\stackrel[\underset{i\neq k}{i=1}]{K}{\sum}p_{i}\, L_{u_{i},b}\varsigma_{i}p_{k}x_{1}+d_{r,b}^{-\alpha_{r}}\sigma_{r}^{2}\nu_{k}p_{k}x_{1},$

$q_{2}=\left(\upsilon_{k}\sigma_{b}^{2}\stackrel[\underset{i\neq k}{i=1}]{K}{\sum}p_{i}y_{1i}+\upsilon_{k}\sigma_{b}^{2}z_{1}\sigma_{r}^{2}+\upsilon_{k}\sigma_{b}^{2}p_{k}x_{1}\right),$

$q_{3}=\stackrel[\underset{i\neq k}{i=1}]{K}{\sum}p_{i}y_{1i}p_{k}\, L_{u_{k},b}\xi_{k}+z_{1}\sigma_{r}^{2}p_{k}\, L_{u_{k},b}\xi_{k}+\stackrel[\underset{i\neq k}{i=1}]{K}{\sum}p_{i}y_{1i}\stackrel[\underset{i\neq k}{i=1}]{K}{\sum}p_{i}\, L_{u_{i},b}\varsigma_{i}+z_{1}\sigma_{r}^{2}\stackrel[\underset{i\neq k}{i=1}]{K}{\sum}p_{i}\, L_{u_{i},b}\varsigma_{i}+\stackrel[\underset{i\neq k}{i=1}]{K}{\sum}p_{i}y_{1i}d_{r,b}^{-\alpha_{r}}\sigma_{r}^{2}\nu_{k}+z_{1}\sigma_{r}^{2}d_{r,b}^{-\alpha_{r}}\sigma_{r}^{2}\nu_{k},$

$q_{4}=\stackrel[\underset{i\neq k}{i=1}]{K}{\sum}p_{i}y_{2i}p_{k}\, L_{u_{k},b}\xi_{k}+\sigma_{e_{j}}^{2}p_{k}\, L_{u_{k},b}\xi_{k}+\stackrel[\underset{i\neq k}{i=1}]{K}{\sum}p_{i}y_{2i}\stackrel[\underset{i\neq k}{i=1}]{K}{\sum}p_{i}\, L_{u_{i},b}\varsigma_{i}+\sigma_{e_{j}}^{2}\stackrel[\underset{i\neq k}{i=1}]{K}{\sum}p_{i}\, L_{u_{i},b}\varsigma_{i}+\stackrel[\underset{i\neq k}{i=1}]{K}{\sum}p_{i}y_{2i}d_{r,b}^{-\alpha_{r}}\sigma_{r}^{2}\nu_{k}+\sigma_{e_{j}}^{2}d_{r,b}^{-\alpha_{r}}\sigma_{r}^{2}\nu_{k},$

$q_{5}=\left(\stackrel[\underset{i\neq k}{i=1}]{K}{\sum}p_{i}y_{1i}\upsilon_{k}\sigma_{b}^{2}+z_{1}\sigma_{r}^{2}\upsilon_{k}\sigma_{b}^{2}\right)$,
$q_{6}=\stackrel[\underset{i\neq k}{i=1}]{K}{\sum}p_{i}y_{2i}\upsilon_{k}\sigma_{b}^{2}+\sigma_{e_{j}}^{2}\upsilon_{k}\sigma_{b}^{2},$

$q_{7}=\stackrel[\underset{i\neq k}{i=1}]{K}{\sum}p_{i}\, L_{u_{i},b}\varsigma_{i}\varrho^{2}2^{r_{s}}p_{k}x_{2}+\varrho^{2}d_{r,b}^{-\alpha_{r}}\sigma_{r}^{2}\nu_{k}2^{r_{s}}p_{k}x_{2}+\stackrel[\underset{i\neq k}{i=1}]{K}{\sum}p_{i}\, L_{u_{i},b}\varsigma_{i}2^{r_{s}}\sigma_{e_{j}}^{2}+d_{r,b}^{-\alpha_{r}}\sigma_{r}^{2}\nu_{k}2^{r_{s}}\sigma_{e_{j}}^{2}+\stackrel[\underset{i\neq k}{i=1}]{K}{\sum}p_{i}\, L_{u_{i},b}\varsigma_{i}2^{r_{s}}\stackrel[\underset{i\neq k}{i=1}]{K}{\sum}p_{i}y_{2i}+d_{r,b}^{-\alpha_{r}}\sigma_{r}^{2}\nu_{k}2^{r_{s}}\stackrel[\underset{i\neq k}{i=1}]{K}{\sum}p_{i}y_{2i},$

$q_{8}=\upsilon_{k}\sigma_{b}^{2}2^{r_{s}}p_{k}x_{2}+\upsilon_{k}\sigma_{b}^{2}2^{r_{s}}\sigma_{e_{j}}^{2}+\upsilon_{k}\sigma_{b}^{2}2^{r_{s}}\stackrel[\underset{i\neq k}{i=1}]{K}{\sum}p_{i}y_{2i}.$

\noindent  Thus, from (\ref{eq:24}), the RIS power should be higher
than or equal to

\begin{equation}
P_{r}=M\left(\frac{-\left(q_{2}-q_{4}-q_{5}+q_{7}\right)\pm\sqrt{\left(q_{2}-q_{4}-q_{5}+q_{7}\right)^{2}-4\left(q_{1}-q_{3}\right)\left(q_{8}-q_{6}\right)}}{2\left(q_{1}-q_{3}\right)}\right)\left(\stackrel[k=1]{K}{\sum}\frac{p_{k}}{d_{u_{k},r}^{\alpha_{r}}}+\sigma_{r}^{2}\right).\label{eq:57}
\end{equation}

\textbf{C)} If user $k$ and the RIS have limited amount of power,
e,g., user $k$ power, $p_{k}$, is less than the required power in
(\ref{eq:49}) and the RIS power, $P_{r}$, is less than the required
power in (\ref{eq:57}). In this case EH RIS can be implemented to
provide the target secrecy rate. Based on (\ref{eq:38}) and (\ref{eq:57}),
the required BS power, $P_{b}$, to charge the RIS and achieve the
target secrecy rate, $r_{s}$, can be obtained by

\[
P_{b}=\frac{M\left(1-\tau\right)}{\eta_{eff}\tau NM}\left(\frac{-\left(q_{2}-q_{4}-q_{5}+q_{7}\right)\pm\sqrt{\left(q_{2}-q_{4}-q_{5}+q_{7}\right)^{2}-4\left(q_{1}-q_{3}\right)\left(q_{8}-q_{6}\right)}}{2\left(q_{1}-q_{3}\right)}\right)
\]

\begin{equation}
\times\left(\stackrel[k=1]{K}{\sum}\frac{p_{k}}{d_{u_{k},r}^{\alpha_{r}}}+\sigma_{r}^{2}\right).
\end{equation}

\section{Numerical Results\label{sec:Numerical-Results}}

In this section, we present simulation and numerical results to assess
the accuracy of the derived expressions and the secrecy performance
of the RIS schemes considered in this paper. Monte-Carlo simulations
with $10^{5}$ independent trials are excuted. The locations of the
BS and the RIS are (0 m, 0 m), (20 m, 20 m), respectively, while the
users are scattered on the corners of a square. Specifically, the
coordinates for the users square are (30 m, 5 m), (35 m, 5 m), (30
m,\textminus 5 m), and (35 m,\textminus 5 m), respectively, while
the eavesdroppers are distributed in a circle centered at (20 m, 0
m) with radius of 10 m.  Unless otherwise specified, the simulation
settings are assumed as follows: $K=J=4$, $N=10$, $M=5$, the users
power $p_{i}=2\textrm{W}$, the active RIS power $P_{r}=7\textrm{W}$
, the BS power in EH RIS scenario $P_{b}=50\textrm{W}$, and the nodes
have same noise variance, $\sigma^{2}=-70\textrm{ dBm}$. In addition,
the path-loss exponent is $2.7$, the Rician factors $\rho=0.5$.
The values of the AoA and AoD of the BS and the RIS are uni-formally
distributed in $(0,2\pi)$, and the concentration parameter of RIS
phase error $\kappa=2$.

\begin{figure}[H]
\noindent \begin{centering}
\subfloat[\label{fig:1a}Secrecy rate versus user, $k$ , power with phase shift
error. ]{\noindent \begin{centering}
\includegraphics[scale=0.6]{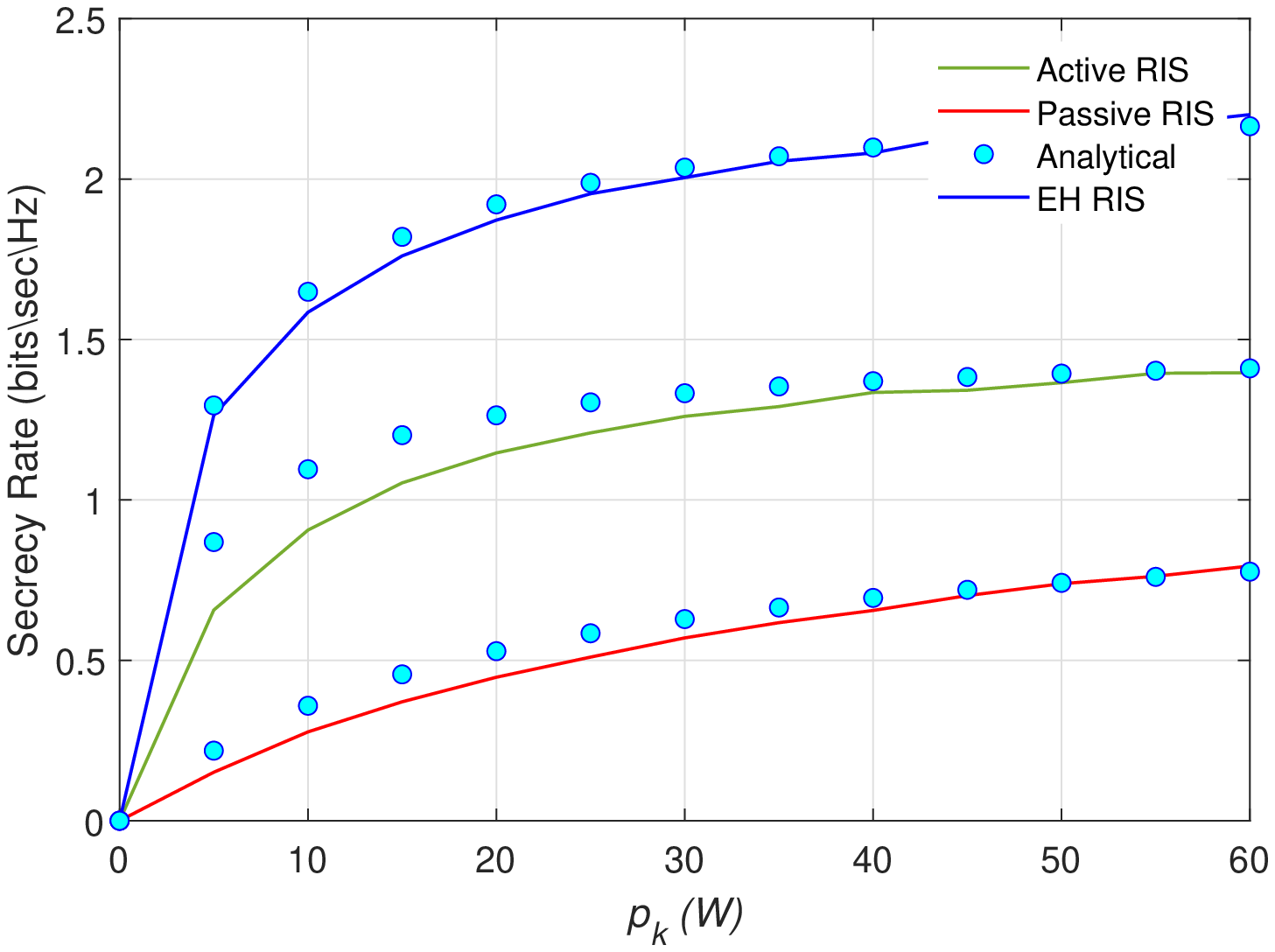}
\par\end{centering}

}\subfloat[\label{fig:1b}Secrecy rate versus user, $k$ , power with no phase
shift error. ]{\noindent \begin{centering}
\includegraphics[scale=0.6]{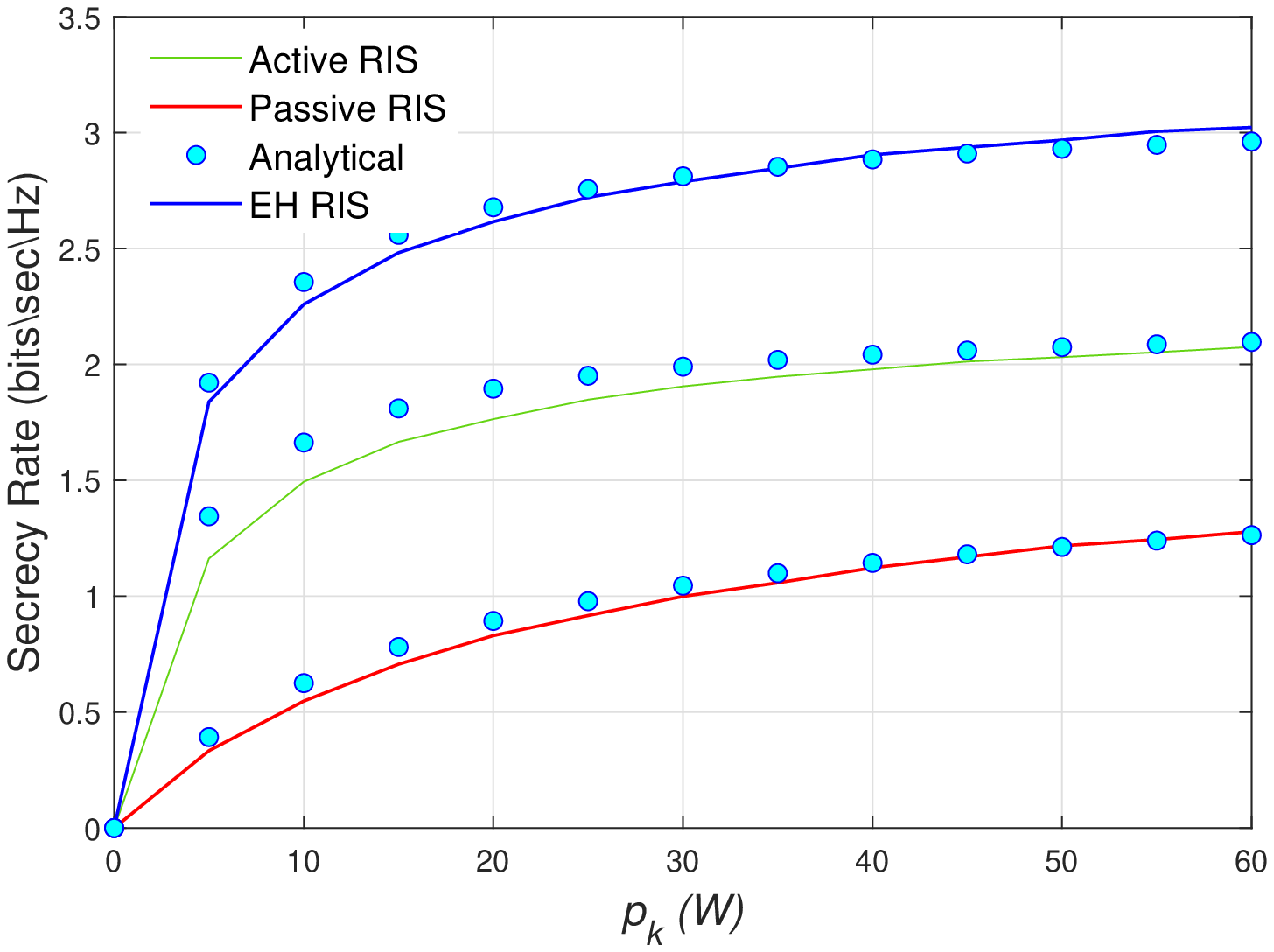}
\par\end{centering}

}
\par\end{centering}

\protect\caption{\label{fig:1}Secrecy rate versus user, $k$ , power with and without
phase shift error. }
\end{figure}

Firstly, in Fig. \ref{fig:1}, we illustrate the ergodic secrecy rate
versus the transmission user power, $p_{k}$, for the three considered
RIS schemes. Fig. \ref{fig:1a} shows the secrecy rate with phase
shift errors and Fig. \ref{fig:1b}, presents the secrecy rate for
the ideal scenario, when there is no phase error at RIS. It is clear
from this figure that the analytical results are in good agreement
with the simulated results, which confirms the validity of the analysis
presented in this paper. It is also evident that for the given parameters
values, the secrecy rate loss due to the imperfect phase shift at
the RIS is about 0.75 bits/s/Hz. In addition, passive RIS achieves
the lowest secrecy rate, but with small amount of power consumption.
The secrecy rate gain of active RIS above passive RIS is about 0.8
bits/s/Hz for a given user power. Furthermore, high secrecy rates
can be achieved and controlled by implementing EH RIS. However, in
this case the BS should transmit high power in the EH phase to provide
sufficient amount of energy at the RIS to achieve higher secrecy rates.

\begin{figure}
\noindent \begin{centering}
\includegraphics[scale=0.7]{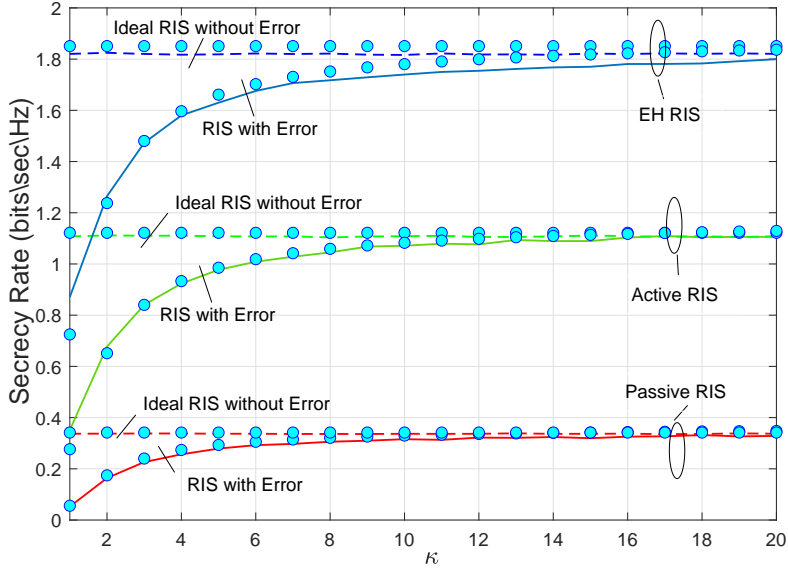}
\par\end{centering}

\protect\caption{\label{fig:2}Secrecy rate versus concentration parameter, $\kappa$,
of RIS phase error.}
\end{figure}

To explain the impact of the phase errors at the RIS on the secrecy
performance, in Fig. \ref{fig:2}, we plot the secrecy rate versus
the concentration parameter of the phase error, $\kappa$. Additionally,
the results of ideal RIS are also presented in this figure. It can
be observed from these results that the secrecy rate enhances as the
concentration parameter, $\kappa$, increases. In addition, at high
concentration parameter values, $\kappa\longrightarrow\infty$, the
secrecy rate achieved by imperfect RIS saturates to that achieved
by ideal RIS. This can be explained by the fact that the phase error
at the RIS is assumed to follow a Von Mises distribution, thus high
concentration parameter values make the error fluctuate in a smaller
range, and when $\kappa\longrightarrow\infty$, the error at the RIS
tends to zero. Accordingly, the secrecy rate of imperfect RIS converges
to the ideal RIS case as $\kappa\longrightarrow\infty$, as explained
in Fig. \ref{fig:2}.

\begin{figure}
\noindent \begin{centering}
\includegraphics[scale=0.7]{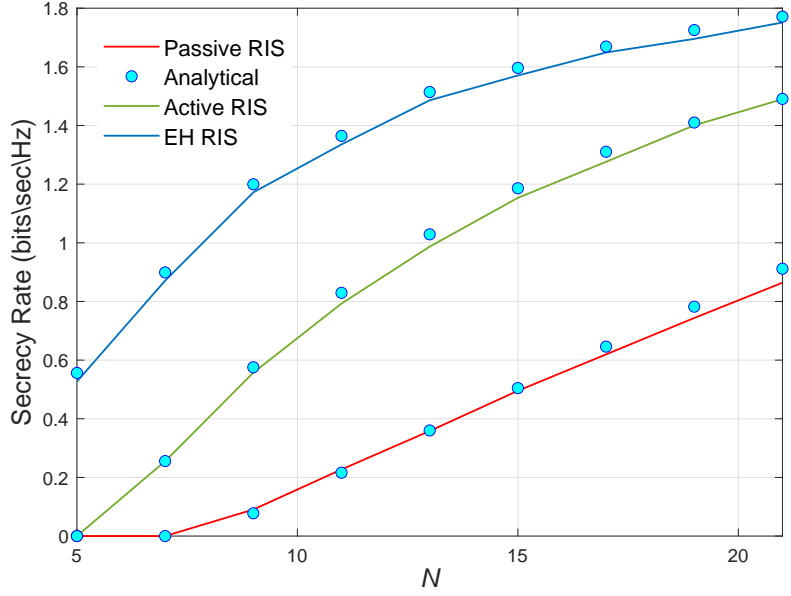}
\par\end{centering}

\protect\caption{\label{fig:3}Secrecy rate versus number of BS antennas, $N$, with
phase shift error. }
\end{figure}

Furthermore, Fig. \ref{fig:3} shows the secrecy rate versus the number
of BS antennas $N$ for the all RIS schemes. It is evident and as
expected, increasing the number of BS antennas $N$ enhances the secrecy
performance for the all RIS schemes. It should be pointed out that
the number of BS antennas, $N$, has impact only on the received signal
at the BS, thus increasing $N$ results in enhancing the rate of the
legitimate users. However $N$ dose not have any impact on the rate
at the eavesdroppers. Having said that in EH RIS, increasing $N$
also increases the amount of the harvested energy at the RIS. Thus,
in EH RIS, $N$ has impact on both achievable rates at the BS and
the eavesdroppers.

\begin{figure}
\noindent \begin{centering}
\includegraphics[scale=0.7]{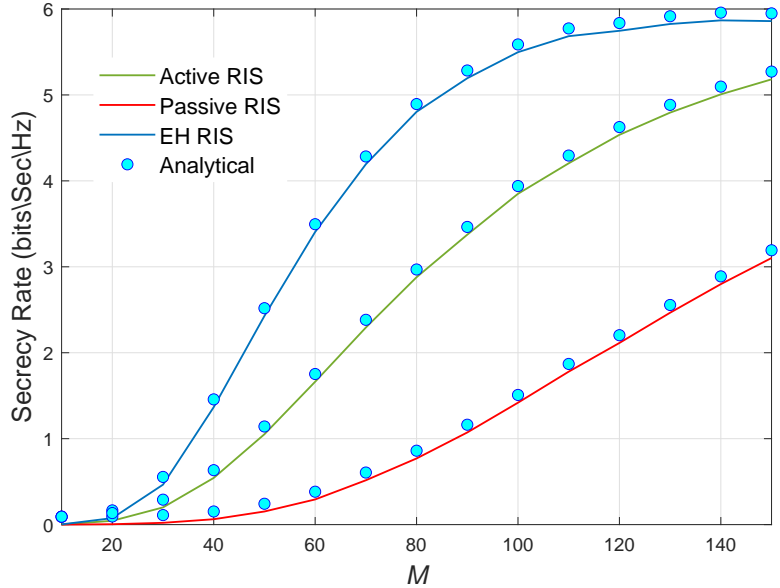}
\par\end{centering}

\protect\caption{\label{fig:4}Secrecy rate versus number of RIS elements, $M$, with
phase shift error. }
\end{figure}

In Fig. \ref{fig:4}, we depict the secrecy rate versus the number
of RIS elements, $M$, for the all considered RIS schemes. To obtain
clear insights and results, in this figure the noise variance at the
nodes is assumed to be $\sigma^{2}=-20\textrm{ dBm}$. Notably and
as expected, increasing $M$ results in enhancing the secrecy rate
for the all considered scenarios. In addition, as we can notice from
the analytical expressions of the secrecy rate presented in this paper,
the number of RIS elements $M$ has impact on both the achievable
rate at the BS and the eavesdroppers, e.g., adding more RIS elements
increases the rate at the BS and the eavesdroppers. However this improvement
in the rate is essential at the BS, because the RIS phase shifts are
designed to be toward the BS direction. Furthermore, in the EH RIS
scheme, increasing the number of the RIS elements, $M$, leads to
an increase in the amount of the harvested energy at the RIS and thus
$P_{r}$ will be high when the number of elements $M$ is very large.

\begin{figure}[H]
\noindent \begin{centering}
\subfloat[\label{fig:a}The user power versus target secrecy rate when $\kappa=2$
.]{\noindent \begin{centering}
\includegraphics[scale=0.6]{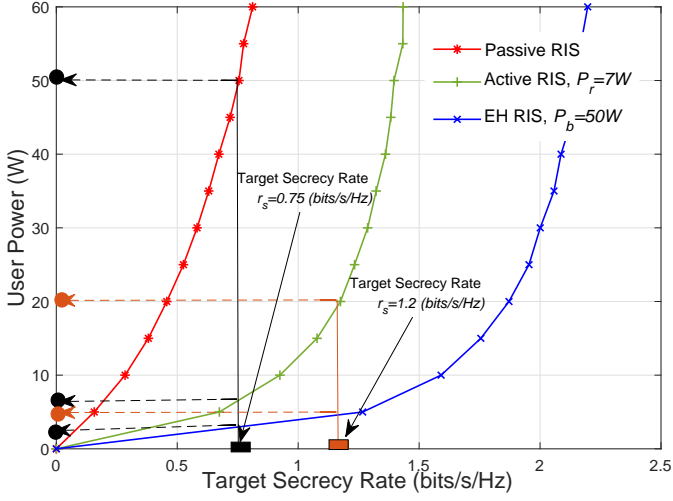}
\par\end{centering}

}\subfloat[\label{fig:b}The user power versus target secrecy rate when $\kappa=8$.]{\noindent \begin{centering}
\includegraphics[scale=0.6]{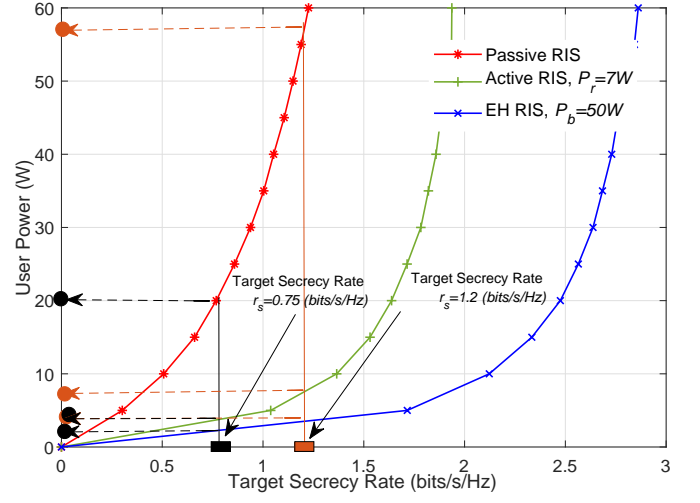}
\par\end{centering}

}
\par\end{centering}

\noindent \begin{centering}
\subfloat[\label{fig:c}RIS configuration selection scheme when $p_{k}=20W$
and $\kappa=2$.]{\noindent \begin{centering}
\includegraphics[scale=0.6]{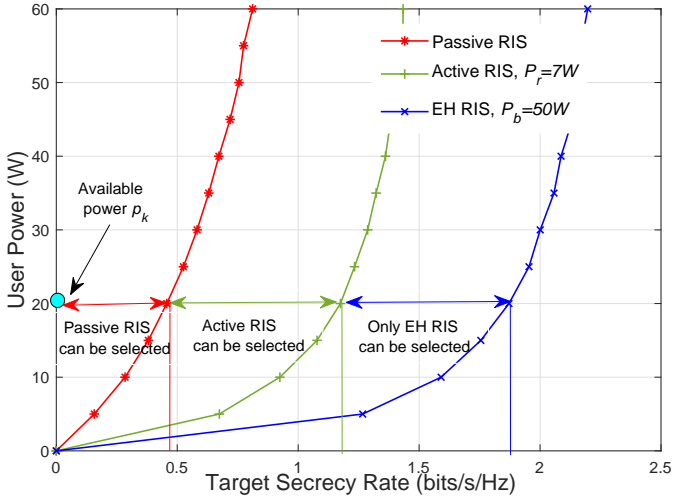}
\par\end{centering}

}\subfloat[\label{fig:d}RIS configuration selection scheme when $p_{k}=20W$
and, $\kappa=8$.]{\noindent \begin{centering}
\includegraphics[scale=0.6]{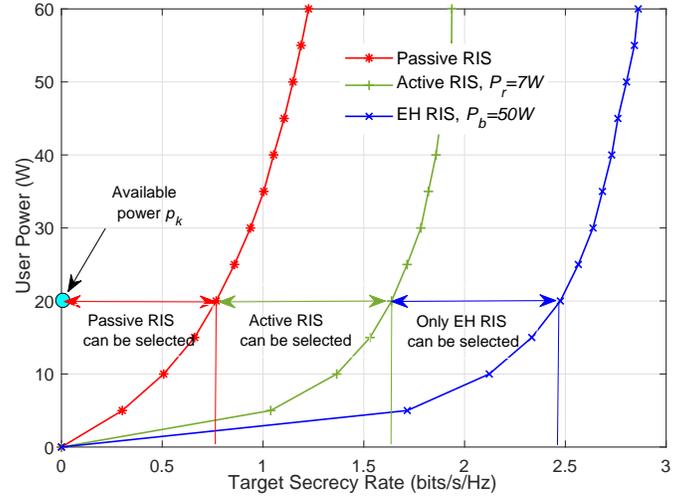}
\par\end{centering}

}
\par\end{centering}

\protect\caption{\label{fig:6}The user power versus target secrecy rate for different
values of the concentration parameter of RIS phase error, $\kappa$.}
\end{figure}

In order to illustrate the RIS configuration selection scheme, in
Fig. \ref{fig:6} we plot the user power versus the target secrecy
rate for different values of the concentration parameter of RIS phase
error, $\kappa=2\textrm{ and }8$. Firstly, in Figs. \ref{fig:a}
and \ref{fig:b}, we consider two examples, when the target secrecy
rate is assumed to be $r_{s}=0.75\,\left(bits/s/Hz\right)$ and $r_{s}=1.2\,\left(bits/s/Hz\right)$
for $\kappa=2\textrm{ and }8$. As we can see from the results in
Fig. \ref{fig:a}, when $r_{s}=0.75\,\left(bits/s/Hz\right)$, passive
RIS can achieve the target secrecy rate with total transmission power
is $P_{T}=p_{k}=50W$, (neglecting the small amount of power consuming
at passive RIS elements), and in the active RIS scheme the user transmission
power can be reduced to around $p_{k}=7W$ and thus the total transmission
power is $P_{T}=p_{k}+P_{r}=14W$, while EH RIS scheme can achieve
the target secrecy rate with the smallest amount of the user power
which is about $p_{k}=2.95W$, but with the highest total transmission
power $P_{T}=p_{k}+P_{b}=52.95W$. Similar observations can be noticed
from the second scenario when $r_{s}=1.2\,\left(bits/s/Hz\right)$,
passive RIS achieves the target secrecy rate with the highest user
power, while EH RIS achieves, $r_{s}$, with the smallest user power
but with very high total consumption power, and the active RIS scheme
works between these two regions. In addition, the concentration parameter
of RIS phase error, $\kappa$, has essential impact on the required
user power. By comparing Figs \ref{fig:a} and \ref{fig:b}, one can
notice that as $\kappa$ increases the required user power to achieve
the target secrecy rate decreases. For instance when the target secrecy
rate is $r_{s}=0.75\,\left(bits/s/Hz\right)$, the required user power
in the passive RIS scheme is about 50W when $\kappa=2$, and 20W when
$\kappa=8$. This is due to the fact explained in Fig. \ref{fig:2}.

Then, in Figs. \ref{fig:c} and \ref{fig:d}, we present the RIS configuration
selection scheme when the available user power is $p_{k}=20W$ for
$\kappa=2\textrm{ and }8$. In the first case when $\kappa=2$, if
the target secrecy rate is $r_{s}\leq0.45\,\left(bits/s/Hz\right)$,
passive RIS can be selected, and active RIS can be implemented if
the target secrecy rate is $r_{s}\leq1.17\,\left(bits/s/Hz\right)$,
while EH RIS can be selected if $r_{s}\leq1.87\,\left(bits/s/Hz\right)$.
These secrecy rate regions of the RIS schemes become wider as the
concentration parameter of RIS phase error, $\kappa$, increases.
In Fig. \ref{fig:d} when $\kappa=8$, passive RIS can be selected
to achieve secrecy rates up to $r_{s}\leq0.77\,\left(bits/s/Hz\right)$,
and active RIS can be selected to perform secrecy rates less than
or equal to $r_{s}\leq1.635\,\left(bits/s/Hz\right)$, whilst EH RIS
can be used to achieve secrecy rates up to $r_{s}\leq2.48\,\left(bits/s/Hz\right)$.

\section{Conclusions\label{sec:Conclusions}}

In this paper the impact of phase shift error on the secrecy performance
of up-link RIS-aided MU-MISO systems was considered. Under Rician
fading channels and phase shift errors the ergodic secrecy rate for,
passive RIS, active RIS, and EH RIS have been analyzed. Then, the
phase shifts at the RIS have been optimized based on the derived rate
expressions. In addition, according to the target secrecy rate and
amount of power available at the users, the best RIS configuration
selection scheme has been considered. The results presented in this
work demonstrated that an active RIS scheme can enhance the secrecy
performance of imperfect RIS elements, especially when the users have
limited amount of power. Furthermore, increasing the number of BS
antennas, the concentration parameter of RIS phase error, and the
number of RIS elements lead to the enhancement of the secrecy performance.

\section*{Appendix A}

By using Jensen inequality, the ergodic rate can be expressed as 

\begin{equation}
\mathscr{E}\left\{ R_{b_{k}}\right\} \approx\log_{2}\left(1+\mathscr{E}\left\{ \frac{p_{k}\, L_{u_{k},{}_{b}}\left|\mathbf{h}_{r,k}^{H}\Theta^{H}\mathbf{G}^{H}\mathbf{G}\Theta\bar{\Theta}\mathbf{h}_{r,k}\right|^{2}}{\stackrel[\underset{i\neq k}{i=1}]{K}{\sum}p_{i}\, L_{u_{i},b}\left|\mathbf{h}_{r,k}^{H}\Theta^{H}\mathbf{G}^{H}\mathbf{G}\Theta\bar{\Theta}\mathbf{h}_{r,i}\right|^{2}+\left\Vert \mathbf{h}_{r,k}^{H}\Theta^{H}\mathbf{G}^{H}\right\Vert ^{2}\sigma_{b}^{2}}\right\} \right).
\end{equation}

Due to the paper length limitation, in this Appendix we will explain
how to calculate the average of the first term, similarly and by following
similar steps we can find the average of the other terms. The first
term is

\begin{equation}
\mathscr{E}\left\{ P_{k}\, L_{u_{k},b}\left|\mathbf{h}_{r,k}^{H}\Theta^{H}\mathbf{G}^{H}\mathbf{G}\Theta\bar{\Theta}\mathbf{h}_{r,k}\right|^{2}\right\} =P_{k}\, L_{u_{k},b}\mathscr{E}\left\{ \left|\mathbf{h}_{r,k}^{H}\Theta^{H}\mathbf{G}^{H}\mathbf{G}\Theta\bar{\Theta}\mathbf{h}_{r,k}\right|^{2}\right\} 
\end{equation}

\noindent where

\[
\mathbf{h}_{r,k}^{H}\Theta^{H}\mathbf{G}^{H}\mathbf{G}\Theta\bar{\Theta}\mathbf{h}_{r,k}=\mathbf{h}_{r,k}^{H}\Theta^{H}\left(\frac{\rho_{b}}{\rho_{b}+1}\mathbf{\bar{G}}^{H}\mathbf{\bar{G}}+\frac{\sqrt{\rho_{b}}}{\rho_{b}+1}\mathbf{\bar{G}}^{H}\mathbf{\tilde{G}}+\frac{\sqrt{\rho_{b}}}{\rho_{b}+1}\mathbf{\tilde{G}}^{H}\mathbf{\bar{G}}+\frac{1}{\rho_{b}+1}\mathbf{\tilde{G}}^{H}\mathbf{\tilde{G}}\right)\Theta\bar{\Theta}\mathbf{h}_{r,k}
\]

\begin{equation}
=\frac{1}{\rho_{b}+1}\mathbf{h}_{r,k}^{H}\Theta^{H}\left(\rho_{b}\mathbf{\bar{G}}^{H}\mathbf{\bar{G}}+\sqrt{\rho_{b}}\mathbf{\bar{G}}^{H}\mathbf{\tilde{G}}+\sqrt{\rho_{b}}\mathbf{\tilde{G}}^{H}\mathbf{\bar{G}}+\mathbf{\tilde{G}}^{H}\mathbf{\tilde{G}}\right)\Theta\bar{\Theta}\mathbf{h}_{r,k}=\frac{1}{\rho_{b}+1}\mathbf{h}_{r,k}^{H}\mathbf{A}\bar{\Theta}\mathbf{h}_{r,k}\label{eq:54}
\end{equation}

\noindent where $\mathbf{A}=\Theta^{H}\left(\rho_{b}\mathbf{\bar{G}}^{H}\mathbf{\bar{G}}+\sqrt{\rho_{b}}\mathbf{\bar{G}}^{H}\mathbf{\tilde{G}}+\sqrt{\rho_{b}}\mathbf{\tilde{G}}^{H}\mathbf{\bar{G}}+\mathbf{\tilde{G}}^{H}\mathbf{\tilde{G}}\right)\Theta$.
Now (\ref{eq:54}) can be expressed as

\[
\mathbf{h}_{r,k}^{H}\Theta^{H}\mathbf{G}^{H}\mathbf{G}\Theta\bar{\Theta}\mathbf{h}_{r,k}=\frac{1}{\left(\rho_{b}+1\right)\left(\rho_{k}+1\right)}\left(\sqrt{\rho_{k}}\mathbf{\bar{h}}_{r,k}^{H}+\mathbf{\tilde{h}}_{r,k}^{H}\right)\mathbf{A}\bar{\Theta}\left(\sqrt{\rho_{k}}\mathbf{\bar{h}}_{r,k}+\mathbf{\tilde{h}}_{r,k}\right)
\]

\begin{equation}
=\frac{1}{\left(\rho_{b}+1\right)\left(\rho_{k}+1\right)}\left(\underset{\Delta_{1}}{\underbrace{\rho_{k}\mathbf{\bar{h}}_{r,k}^{H}\mathbf{A}\bar{\Theta}\mathbf{\bar{h}}_{r,k}}}+\underset{\Delta_{2}}{\underbrace{\sqrt{\rho_{k}}\mathbf{\bar{h}}_{r,k}^{H}\mathbf{A}\bar{\Theta}\mathbf{\tilde{h}}_{r,k}}}+\underset{\Delta_{3}}{\underbrace{\sqrt{\rho_{k}}\mathbf{\tilde{h}}_{r,k}^{H}\mathbf{A}\bar{\Theta}\mathbf{\bar{h}}_{r,k}}}+\underset{\Delta_{4}}{\underbrace{\mathbf{\tilde{h}}_{r,k}^{H}\mathbf{A}\bar{\Theta}\mathbf{\tilde{h}}_{r,k}}}\right)
\end{equation}

The channels are independent and have zero mean. Thus by removing
the zero expectation terms, we can get

\[
\mathscr{E}\left\{ \left|\mathbf{h}_{r,k}^{H}\Theta^{H}\mathbf{G}^{H}\mathbf{G}\Theta\bar{\Theta}\mathbf{h}_{r,k}\right|^{2}\right\} =\frac{1}{\left(\rho_{b}+1\right)^{2}\left(\rho_{k}+1\right)^{2}}\mathscr{E}\left\{ \left|\stackrel[i=1]{4}{\sum}\Delta_{i}\right|^{2}\right\} 
\]

\begin{equation}
=\frac{1}{\left(\rho_{b}+1\right)^{2}\left(\rho_{k}+1\right)^{2}}\left(\stackrel[i=1]{4}{\sum}\mathscr{E}\left\{ \left|\Delta_{i}\right|^{2}\right\} +2\mathscr{E}\left\{ \Delta_{1}\Delta_{4}^{*}\right\} \right)
\end{equation}

\noindent Now the first term

\[
\Delta_{1}=\rho_{k}\mathbf{\bar{h}}_{r,k}^{H}\Theta^{H}\left(\rho_{b}\mathbf{\bar{G}}^{H}\mathbf{\bar{G}}+\sqrt{\rho_{b}}\mathbf{\bar{G}}^{H}\mathbf{\tilde{G}}+\sqrt{\rho_{b}}\mathbf{\tilde{G}}^{H}\mathbf{\bar{G}}+\mathbf{\tilde{G}}^{H}\mathbf{\tilde{G}}\right)\Theta\bar{\Theta}\mathbf{\bar{h}}_{r,k}
\]

\[
=\left(\underset{\Delta_{1,1}}{\underbrace{\rho_{b}\rho_{k}\mathbf{\bar{h}}_{r,k}^{H}\Theta^{H}\mathbf{\bar{G}}^{H}\mathbf{\bar{G}}\Theta\bar{\Theta}\mathbf{\bar{h}}_{r,k}}}+\underset{\Delta_{1,2}}{\underbrace{\sqrt{\rho_{b}}\rho_{k}\mathbf{\bar{h}}_{r,k}^{H}\Theta^{H}\mathbf{\bar{G}}^{H}\mathbf{\tilde{G}}\Theta\bar{\Theta}\mathbf{\bar{h}}_{r,k}}}\right.
\]

\begin{equation}
\left.+\underset{\Delta_{1,3}}{\underbrace{\sqrt{\rho_{b}}\rho_{k}\mathbf{\bar{h}}_{r,k}^{H}\Theta^{H}\mathbf{\tilde{G}}^{H}\mathbf{\bar{G}}\Theta\bar{\Theta}\mathbf{\bar{h}}_{r,k}}}+\underset{\Delta_{1,4}}{\underbrace{\rho_{k}\mathbf{\bar{h}}_{r,k}^{H}\Theta^{H}\mathbf{\tilde{G}}^{H}\mathbf{\tilde{G}}\Theta\bar{\Theta}\mathbf{\bar{h}}_{r,k}}}\right)
\end{equation}

\noindent The average of the first term

\begin{equation}
\mathscr{E}\left\{ \left|\Delta_{1}\right|^{2}\right\} =\mathscr{E}\left\{ \left|\Delta_{1,1}\right|^{2}\right\} +\mathscr{E}\left\{ \left|\Delta_{1,2}\right|^{2}\right\} +\mathscr{E}\left\{ \left|\Delta_{1,3}\right|^{2}\right\} +\mathscr{E}\left\{ \left|\Delta_{1,4}\right|^{2}\right\} +2\mathscr{E}\left\{ \Delta_{1,1}\Delta_{1,4}^{H}\right\} 
\end{equation}

\noindent where $\Delta_{1,1}=\rho_{b}\rho_{k}\mathbf{\bar{h}}_{r,k}^{H}\Theta^{H}\mathbf{\bar{G}}^{H}\mathbf{\bar{G}}\Theta\bar{\Theta}\mathbf{\bar{h}}_{r,k}$,
 which can be written as

\[
\Delta_{1,1}=\rho_{b}\rho_{k}\mathbf{a}_{M}^{H}\left(\phi_{kr}^{a},\phi_{kr}^{e}\right)\Theta^{H}\mathbf{a}_{M}^{H}\left(\phi_{r}^{a},\phi_{r}^{e}\right)\mathbf{a}_{M}\left(\phi_{r}^{a},\phi_{r}^{e}\right)\Theta\bar{\Theta}\mathbf{a}_{M}\left(\phi_{kr}^{a},\phi_{kr}^{e}\right),
\]

\begin{equation}
\Delta_{1,1}=\rho_{b}\rho_{k}\left(\stackrel[m=1]{M}{\sum}a_{M,m}^{H}\left(\phi_{kr}^{a},\phi_{kr}^{e}\right)e^{-j\varphi_{m}}a_{M,m}^{H}\left(\phi_{r}^{a},\phi_{r}^{e}\right)\right)\left(\stackrel[m=1]{M}{\sum}a_{M,m}\left(\phi_{kr}^{a},\phi_{kr}^{e}\right)e^{j\varphi_{m}}e^{j\bar{\varphi_{m}}}a_{M,m}\left(\phi_{r}^{a},\phi_{r}^{e}\right)\right).
\end{equation}

\noindent The average can now be written as

\[
\mathscr{E}\left\{ \left|\Delta_{1,1}\right|^{2}\right\} =\rho_{b}^{2}\rho_{k}^{2}\left(\stackrel[m=1]{M}{\sum}a_{M,m}^{H}\left(\phi_{kr}^{a},\phi_{kr}^{e}\right)e^{-j\varphi_{m}}a_{M,m}^{H}\left(\phi_{r}^{a},\phi_{r}^{e}\right)\right)^{2}
\]

\begin{equation}
\times\left(M+\rho\left(\kappa\right)^{2}\left|\stackrel[m_{1}=1]{M}{\sum}\stackrel[m_{2}\neq m_{1}]{M}{\sum}\left(a_{M,m_{1}}\left(\phi_{kr}^{a},\phi_{kr}^{e}\right)e^{j\varphi_{m_{1}}}a_{M,m_{1}}\left(\phi_{r}^{a},\phi_{r}^{e}\right)\right)\left(a_{M,m_{2}}\left(\phi_{kr}^{a},\phi_{kr}^{e}\right)e^{j\varphi_{m_{2}}}a_{M,m_{2}}\left(\phi_{r}^{a},\phi_{r}^{e}\right)\right)^{H}\right|^{2}\right)
\end{equation}

\begin{equation}
\mathscr{E}\left\{ \left|\Delta_{1,1}\right|^{2}\right\} =\rho_{b}^{2}\rho_{k}^{2}\left|f_{k}\right|^{2}\left(\left(1-\rho\left(\kappa\right)^{2}\right)M+\rho\left(\kappa\right)^{2}\left|f_{k}\right|^{2}\right)
\end{equation}

\noindent where $f_{k}=\stackrel[m=1]{M}{\sum}f_{k,m},f_{k,m}=a_{M.m}^{H}\left(\phi_{r}^{a},\phi_{r}^{e}\right)e^{j\varphi_{m}}a_{M,m}\left(\phi_{kr}^{a},\phi_{kr}^{e}\right)$.
The second term,

\[
\Delta_{1,2}=\sqrt{\rho_{b}}\rho_{k}a_{M}^{H}\left(\phi_{kr}^{a},\phi_{kr}^{e}\right)\Theta^{H}a_{M}\left(\phi_{r}^{a},\phi_{r}^{e}\right)a_{N}^{H}\left(\phi_{b}^{a},\phi_{b}^{e}\right)\mathbf{\tilde{G}}\Theta\bar{\Theta}a_{M}\left(\phi_{kr}^{a},\phi_{kr}^{e}\right)
\]

\begin{equation}
=\sqrt{\rho_{b}}\rho_{k}f_{k}^{*}\stackrel[m=1]{M}{\sum}\stackrel[n=1]{N}{\sum}a_{N,n}^{H}\left(\phi_{b}^{a},\phi_{b}^{e}\right)\tilde{\mathbf{g}}_{nm}e^{j\varphi_{m}}e^{j\bar{\varphi_{m}}}a_{M,m}\left(\phi_{kr}^{a},\phi_{kr}^{e}\right),
\end{equation}

\begin{equation}
\mathscr{E}\left\{ \left|\Delta_{1,2}\right|^{2}\right\} =\rho_{b}\rho_{k}^{2}NM\left|f_{k}\right|^{2}.
\end{equation}

\noindent The third term

\[
\Delta_{1,3}=\sqrt{\rho_{b}}\rho_{k}a_{M}^{H}\left(\phi_{kr}^{a},\phi_{kr}^{e}\right)\Theta^{H}\mathbf{\tilde{G}}^{H}a_{N}\left(\phi_{b}^{a},\phi_{b}^{e}\right)a_{M}^{H}\left(\phi_{r}^{a},\phi_{r}^{e}\right)\Theta\bar{\Theta}a_{M}\left(\phi_{kr}^{a},\phi_{kr}^{e}\right)
\]

\begin{equation}
=\sqrt{\rho_{b}}\rho_{k}\stackrel[m=1]{M}{\sum}\stackrel[n=1]{N}{\sum}a_{N,n}^{H}\left(\phi_{b}^{a},\phi_{b}^{e}\right)\tilde{g}_{nm}^{H}e^{-j\varphi_{m}}a_{M,m}\left(\phi_{kr}^{a},\phi_{kr}^{e}\right)\stackrel[m=1]{M}{\sum}e^{j\bar{\varphi_{m}}}f_{k,m},
\end{equation}

\begin{equation}
\mathscr{E}\left\{ \left|\Delta_{1,3}\right|^{2}\right\} =\rho_{b}\rho_{k}^{2}\left(NM\rho\left(\kappa\right)^{2}\left|f_{k}\right|^{2}+\left(1-\rho\left(\kappa\right)^{2}\right)NM^{2}\right).
\end{equation}

\noindent The forth term

\[
\Delta_{1,4}=\rho_{k}a_{M}^{H}\left(\phi_{kr}^{a},\phi_{kr}^{e}\right)\Theta^{H}\mathbf{\tilde{G}}^{H}\mathbf{\tilde{G}}\Theta\bar{\Theta}a_{M}\left(\phi_{kr}^{a},\phi_{kr}^{e}\right)
\]

\begin{equation}
=\rho_{k}\stackrel[m_{1}=1]{M}{\sum}a_{M,m_{1}}^{H}\left(\phi_{kr}^{a},\phi_{kr}^{e}\right)e^{-j\varphi_{m}}\tilde{g}_{nm_{1}}^{H}\stackrel[m_{2}=1]{M}{\sum}\tilde{g}_{nm_{2}}e^{j\varphi_{m}}e^{j\bar{\varphi_{m}}}a_{M,m_{2}}\left(\phi_{kr}^{a},\phi_{kr}^{e}\right),
\end{equation}

\begin{equation}
\mathscr{E}\left\{ \left|\Delta_{1,4}\right|^{2}\right\} =\rho_{k}NM\left(M\rho\left(\kappa\right)^{2}+1-\rho\left(\kappa\right)^{2}\right)+NM^{2}.
\end{equation}

\noindent The last term

\begin{equation}
\mathscr{E}\left\{ \Delta_{1,1}\Delta_{1,4}^{*}\right\} =N\left|f_{k}\right|^{2}\left(M\rho\left(\kappa\right)^{2}+1-\rho\left(\kappa\right)^{2}\right).
\end{equation}

\noindent Similarly, following the same way we can find the average
of the other terms.

\section*{Appendix B}

Using Jensen inequality, the ergodic rate can be written as 

\begin{equation}
\mathscr{E}\left\{ R_{e_{j,k}}\right\} \approx\log_{2}\left(1+\mathscr{E}\left\{ \frac{p_{k}\,\left|d_{u_{k},r}^{-\frac{\alpha_{r}}{2}}d_{e_{j},r}^{-\frac{\alpha_{e}}{2}}\mathbf{h}_{e_{j},r}\Theta\bar{\Theta}\mathbf{h}_{r,k}+\, d_{e_{j},k}^{-\frac{\alpha_{e}}{2}}h_{e_{j},k}\right|^{2}}{\stackrel[\underset{i\neq k}{i=1}]{K}{\sum}p_{i}\,\left|d_{u_{i},r}^{-\frac{\alpha_{r}}{2}}d_{e_{j},r}^{-\frac{\alpha_{e}}{2}}\mathbf{h}_{e_{j},r}\Theta\bar{\Theta}\mathbf{h}_{r,i}+d_{e_{j},i}^{-\frac{\alpha_{e}}{2}}h_{e_{j},i}\right|^{2}+\sigma_{e_{j}}^{2}}\right\} \right)
\end{equation}

The average of the first term, after removing the zero expectation
terms can be calculated by,

\begin{equation}
\mathscr{E}\left\{ \left|d_{u_{k},r}^{-\frac{\alpha_{r}}{2}}d_{e_{j},r}^{-\frac{\alpha_{e}}{2}}\mathbf{h}_{e_{j},r}\Theta\bar{\Theta}\mathbf{h}_{r,k}+\, d_{e_{j},k}^{-\frac{\alpha_{e}}{2}}h_{e_{j},k}\right|^{2}\right\} =d_{u_{k},r}^{-\alpha_{r}}d_{e_{j},r}^{-\alpha_{e}}\mathscr{E}\left\{ \left|\mathbf{h}_{e_{j},r}\Theta\bar{\Theta}\mathbf{h}_{r,k}\right|^{2}\right\} +d_{e_{j},r}^{-\alpha_{e}}
\end{equation}

where 

\[
\mathscr{E}\left\{ \left|\mathbf{h}_{e_{j},r}\Theta\bar{\Theta}\mathbf{h}_{r,k}\right|^{2}\right\} =\mathscr{E}\left\{ \left|\left(\sqrt{\frac{\rho_{e_{j}}}{\rho_{e_{j}}+1}}\sqrt{\frac{\rho_{k}}{\rho_{k}+1}}\mathbf{\bar{h}}_{e_{j}}\Theta\bar{\Theta}\mathbf{\bar{h}}_{r,k}+\sqrt{\frac{\rho_{e_{j}}}{\rho_{e_{j}}+1}}\sqrt{\frac{1}{\rho_{k}+1}}\mathbf{\bar{h}}_{e_{j}}\Theta\bar{\Theta}\mathbf{\tilde{h}}_{r,k}\right.\right.\right.
\]

\begin{equation}
\left.\left.\left.+\sqrt{\frac{\rho_{k}}{\rho_{k}+1}}\sqrt{\frac{1}{\rho_{e_{j}}+1}}\mathbf{\tilde{h}}_{e_{j}}\Theta\bar{\Theta}\mathbf{\bar{h}}_{r,k}+\sqrt{\frac{1}{\rho_{e_{j}}+1}}\sqrt{\frac{1}{\rho_{k}+1}}\mathbf{\tilde{h}}_{e_{j}}\Theta\bar{\Theta}\mathbf{\tilde{h}}_{r,k}\right)\right|^{2}\right\} 
\end{equation}

\[
\mathscr{E}\left\{ \left|\mathbf{h}_{e_{j},r}\Theta\bar{\Theta}\mathbf{h}_{r,k}\right|^{2}\right\} =\frac{\rho_{e_{j}}}{\rho_{e_{j}}+1}\frac{\rho_{k}}{\rho_{k}+1}\mathscr{E}\left|\mathbf{\bar{h}}_{e_{j}}\Theta\bar{\Theta}\mathbf{\bar{h}}_{r,k}\right|^{2}+\frac{\rho_{e_{j}}}{\rho_{e_{j}}+1}\frac{1}{\rho_{k}+1}\mathscr{E}\left|\mathbf{\bar{h}}_{e_{j}}\Theta\bar{\Theta}\mathbf{\tilde{h}}_{r,k}\right|^{2}
\]

\begin{equation}
+\frac{\rho_{k}}{\rho_{k}+1}\frac{1}{\rho_{e_{j}}+1}\mathscr{E}\left|\mathbf{\tilde{h}}_{e_{j}}\Theta\bar{\Theta}\mathbf{\bar{h}}_{r,k}\right|^{2}+\frac{1}{\rho_{e_{j}}+1}\frac{1}{\rho_{k}+1}\mathscr{E}\left|\mathbf{\tilde{h}}_{e_{j}}\Theta\bar{\Theta}\mathbf{\tilde{h}}_{r,k}\right|^{2}
\end{equation}

Now

\begin{equation}
\mathbf{\bar{h}}_{e_{j}}\Theta\bar{\Theta}\mathbf{\bar{h}}_{r,k}=\left(\stackrel[m=1]{M}{\sum}a_{M,m}\left(\phi_{kr}^{a},\phi_{kr}^{e}\right)e^{j\varphi_{m}}e^{j\bar{\varphi_{m}}}a_{M,m}\left(\phi_{e_{j}r}^{a},\phi_{e_{j}r}^{e}\right)\right)
\end{equation}

\[
\mathscr{E}\left|\mathbf{\bar{h}}_{e_{j}}\Theta\bar{\Theta}\mathbf{\bar{h}}_{r,k}\right|^{2}=\mathscr{E}\left|\stackrel[m=1]{M}{\sum}a_{M,m}\left(\phi_{kr}^{a},\phi_{kr}^{e}\right)e^{j\varphi_{m}}e^{j\bar{\varphi_{m}}}a_{M,m}\left(\phi_{e_{j}r}^{a},\phi_{e_{j}r}^{e}\right)\right|^{2}
\]

\[
=M+\rho\left(\kappa\right)^{2}\stackrel[m_{1}=1]{M}{\sum}\stackrel[m_{2}\neq m_{1}]{M}{\sum}\left(a_{M,m_{1}}\left(\phi_{kr}^{a},\phi_{kr}^{e}\right)e^{j\varphi_{m_{1}}}a_{M,m_{1}}\left(\phi_{r}^{a},\phi_{r}^{e}\right)\right)\left(a_{M,m_{2}}\left(\phi_{kr}^{a},\phi_{kr}^{e}\right)e^{j\varphi_{m_{2}}}a_{M,m_{2}}\left(\phi_{r}^{a},\phi_{r}^{e}\right)\right)^{H}
\]

\begin{equation}
=M+\rho\left(\kappa\right)^{2}\xi
\end{equation}

where $\xi=\stackrel[m_{1}=1]{M}{\sum}\stackrel[m_{2}\neq m_{1}]{M}{\sum}\left(a_{M,m_{1}}\left(\phi_{kr}^{a},\phi_{kr}^{e}\right)e^{j\varphi_{m_{1}}}a_{M,m_{1}}\left(\phi_{r}^{a},\phi_{r}^{e}\right)\right)\left(a_{M,m_{2}}\left(\phi_{kr}^{a},\phi_{kr}^{e}\right)e^{j\varphi_{m_{2}}}a_{M,m_{2}}\left(\phi_{r}^{a},\phi_{r}^{e}\right)\right)^{H}$.
Similarly, the second term,

\begin{equation}
\mathbf{\bar{h}}_{e_{j}}\Theta\bar{\Theta}\mathbf{\tilde{h}}_{r,k}=\mathbf{a}_{M}\left(\phi_{e_{j}r}^{a},\phi_{e_{j}r}^{e}\right)\Theta\bar{\Theta}\mathbf{\tilde{h}}_{r,k}=\stackrel[m=1]{M}{\sum}a_{Mm}\left(\phi_{e_{j}r}^{a},\phi_{e_{j}r}^{e}\right)e^{j\varphi_{m}}e^{j\bar{\varphi_{m}}}\left[\tilde{h}_{r,k}\right]_{m}
\end{equation}

\[
\mathscr{E}\left|\mathbf{\bar{h}}_{e_{j}}\Theta\bar{\Theta}\mathbf{\tilde{h}}_{r,k}\right|^{2}=\mathscr{E}\left|\stackrel[m=1]{M}{\sum}a_{Mm}\left(\phi_{e_{j}r}^{a},\phi_{e_{j}r}^{e}\right)e^{j\varphi_{m}}e^{j\bar{\varphi_{m}}}\left[\tilde{h}_{r,k}\right]_{m}\right|^{2}
\]

\[
\mathscr{E}\left|\mathbf{\bar{h}}_{e_{j}}\Theta\bar{\Theta}\mathbf{\tilde{h}}_{r,k}\right|^{2}=M+
\]

\begin{equation}
\mathscr{E}\left\{ \stackrel[m_{1}=1]{M}{\sum}\stackrel[m_{2}\neq m_{1}]{M}{\sum}\left(a_{Mm_{1}}\left(\phi_{e_{j}r}^{a},\phi_{e_{j}r}^{e}\right)e^{j\varphi_{m_{1}}}e^{j\bar{\varphi_{m1}}}\left[\tilde{h}_{r,k}\right]_{m_{1}}\right)\left(a_{Mm2}\left(\phi_{e_{j}r}^{a},\phi_{e_{j}r}^{e}\right)e^{j\varphi_{m2}}e^{j\bar{\varphi_{m2}}}\left[\tilde{h}_{r,k}\right]_{m_{2}}\right)^{H}\right\} =M
\end{equation}

other terms,

\begin{equation}
\mathbf{\tilde{h}}_{e_{j}}\Theta\bar{\Theta}\mathbf{\bar{h}}_{r,k}=\stackrel[m=1]{M}{\sum}\tilde{h}_{e_{j,m}}e^{j\varphi_{m}}e^{j\bar{\varphi_{m}}}a_{M,m}\left(\phi_{kr}^{a},\phi_{kr}^{e}\right)
\end{equation}

\[
\mathscr{E}\left|\mathbf{\tilde{h}}_{e_{j}}\Theta\bar{\Theta}\mathbf{\bar{h}}_{r,k}\right|^{2}=M+
\]

\begin{equation}
\mathscr{E}\left\{ \stackrel[m_{1}=1]{M}{\sum}\stackrel[m_{2}\neq m_{1}]{M}{\sum}\left(\left[\tilde{h}_{e_{j}}\right]_{m_{1}}e^{j\varphi_{m_{1}}}e^{j\bar{\varphi_{m1}}}a_{Mm1}\left(\phi_{kr}^{a},\phi_{kr}^{e}\right)\right)\left(\left[\tilde{h}_{e_{j}}\right]_{m_{2}}e^{j\varphi_{m_{2}}}e^{j\bar{\varphi_{m2}}}a_{Mm2}\left(\phi_{kr}^{a},\phi_{kr}^{e}\right)\right)^{H}\right\} =M
\end{equation}

and

\begin{equation}
\mathbf{\tilde{h}}_{e_{j}}\Theta\bar{\Theta}\mathbf{\tilde{h}}_{r,k}=\stackrel[m=1]{M}{\sum}\left[\tilde{h}_{e_{j}}\right]_{m}e^{j\varphi_{m}}e^{j\bar{\varphi_{m}}}\left[\tilde{h}_{r,k}\right]_{m}
\end{equation}

\begin{equation}
\mathscr{E}\left|\mathbf{\tilde{h}}_{e_{j}}\Theta\bar{\Theta}\mathbf{\tilde{h}}_{r,k}\right|^{2}=\mathscr{E}\left|\stackrel[m=1]{M}{\sum}\left[\tilde{h}_{e_{j}}\right]_{m}e^{j\varphi_{m}}e^{j\bar{\varphi_{m}}}\left[\tilde{h}_{r,k}\right]_{m}\right|^{2}=M
\end{equation}

Now, we are ready to write the average of the first term, 

\[
\mathscr{E}\left\{ \left|\mathbf{h}_{e_{j},r}\Theta\bar{\Theta}\mathbf{h}_{r,k}\right|^{2}\right\} =\frac{\rho_{e_{j}}}{\rho_{e_{j}}+1}\frac{\rho_{k}}{\rho_{k}+1}\left(M+\rho\left(\kappa\right)^{2}\xi\right)+\frac{\rho_{e_{j}}}{\rho_{e_{j}}+1}\frac{1}{\rho_{k}+1}M
\]

\begin{equation}
+\frac{\rho_{k}}{\rho_{k}+1}\frac{1}{\rho_{e_{j}}+1}M+\frac{1}{\rho_{e_{j}}+1}\frac{1}{\rho_{k}+1}M
\end{equation}

Similarly we can find the average of the second term as, 

\begin{equation}
\mathscr{E}\left\{ \left|d_{u_{i},r}^{-\frac{\alpha_{r}}{2}}d_{e_{j},r}^{-\frac{\alpha_{e}}{2}}\mathbf{h}_{e_{j},r}\Theta\bar{\Theta}\mathbf{h}_{r,i}+d_{e_{j},i}^{-\frac{\alpha_{e}}{2}}h_{e_{j},i}\right|^{2}\right\} =d_{u_{i},r}^{-\alpha_{r}}d_{e_{j},r}^{-\alpha_{e}}\mathscr{E}\left\{ \left|\mathbf{h}_{e_{j},r}\Theta\bar{\Theta}\mathbf{h}_{r,i}\right|^{2}\right\} +d_{e_{j},i}^{-\alpha_{e}}
\end{equation}

\[
\mathscr{E}\left\{ \left|\mathbf{h}_{e_{j},r}\Theta\bar{\Theta}\mathbf{h}_{r,i}\right|^{2}\right\} =\frac{\rho_{e_{j}}}{\rho_{e_{j}}+1}\frac{\rho_{i}}{\rho_{i}+1}\left(M+\rho\left(\kappa\right)^{2}\xi\right)+\frac{\rho_{e_{j}}}{\rho_{e_{j}}+1}\frac{1}{\rho_{i}+1}M
\]

\begin{equation}
+\frac{\rho_{i}}{\rho_{i}+1}\frac{1}{\rho_{e_{j}}+1}M+\frac{1}{\rho_{e_{j}}+1}\frac{1}{\rho_{i}+1}M
\end{equation}

\section*{Appendix C}

Using Jensen inequality, the ergodic rate can be written as 

\begin{equation}
\mathscr{E}\left\{ R_{b_{k}}\right\} \approx\log_{2}\left(1+\mathscr{E}\left\{ \gamma_{b_{k}}\right\} \right)
\end{equation}

We will follow similar steps as in Appendix A,

\begin{equation}
\mathbf{h}_{r,k}^{H}\Theta^{H}\mathbf{G}^{H}\mathbf{G}\bar{\Theta}\Theta=\frac{1}{\rho_{b}+1}\mathbf{h}_{r,k}^{H}\mathbf{A}\bar{\Theta}
\end{equation}

where $\mathbf{A}=\Theta^{H}\left(\rho_{b}\mathbf{\bar{G}}^{H}\mathbf{\bar{G}}+\sqrt{\rho_{b}}\mathbf{\bar{G}}^{H}\mathbf{\tilde{G}}+\sqrt{\rho_{b}}\mathbf{\tilde{G}}^{H}\mathbf{\bar{G}}+\mathbf{\tilde{G}}^{H}\mathbf{\tilde{G}}\right)\Theta.$
Last expression can be written as

\[
\mathbf{h}_{r,k}^{H}\Theta^{H}\mathbf{G}^{H}\mathbf{G}\bar{\Theta}\Theta=\frac{1}{\left(\rho_{b}+1\right)\sqrt{\left(\rho_{k}+1\right)}}\left(\sqrt{\rho_{k}}\mathbf{\bar{h}}_{r,k}^{H}+\mathbf{\tilde{h}}_{r,k}^{H}\right)\mathbf{A}\bar{\Theta}
\]

\begin{equation}
=\frac{1}{\left(\rho_{b}+1\right)\sqrt{\left(\rho_{k}+1\right)}}\left(\underset{\Delta_{1}}{\underbrace{\sqrt{\rho_{k}}\mathbf{\bar{h}}_{r,k}^{H}\mathbf{A}\bar{\Theta}}}+\underset{\Delta_{2}}{\underbrace{\mathbf{\tilde{h}}_{r,k}^{H}\mathbf{A}\bar{\Theta}}}\right)
\end{equation}

The average can be written as,

\begin{equation}
\mathscr{E}\left\{ \left\Vert \mathbf{h}_{r,k}^{H}\Theta^{H}\mathbf{G}^{H}\mathbf{G}\bar{\Theta}\Theta\right\Vert ^{2}\right\} =\frac{1}{\left(\rho_{b}+1\right)^{2}\left(\rho_{k}+1\right)}\mathscr{E}\left\{ \rho_{r}\left\Vert \underset{\Delta_{1}}{\underbrace{\mathbf{\bar{h}}_{r,k}^{H}\mathbf{A}\bar{\Theta}}}\right\Vert ^{2}+\left\Vert \underset{\Delta_{2}}{\underbrace{\mathbf{\tilde{h}}_{r,k}^{H}\mathbf{A}\bar{\Theta}}}\right\Vert ^{2}\right\} 
\end{equation}

\[
\Delta_{1}=\sqrt{\rho_{k}}\mathbf{\bar{h}}_{r,k}^{H}\Theta^{H}\left(\rho_{b}\mathbf{\bar{G}}^{H}\mathbf{\bar{G}}+\sqrt{\rho_{b}}\mathbf{\bar{G}}^{H}\mathbf{\tilde{G}}+\sqrt{\rho_{b}}\mathbf{\tilde{G}}^{H}\mathbf{\bar{G}}+\mathbf{\tilde{G}}^{H}\mathbf{\tilde{G}}\right)\Theta\bar{\Theta}
\]

\[
=\left(\underset{\Delta_{1,1}}{\underbrace{\rho_{b}\sqrt{\rho_{k}}\mathbf{\bar{h}}_{r,k}^{H}\Theta^{H}\mathbf{\bar{G}}^{H}\mathbf{\bar{G}}\Theta\bar{\Theta}}}+\underset{\Delta_{1,2}}{\underbrace{\sqrt{\rho_{b}}\sqrt{\rho_{k}}\mathbf{\bar{h}}_{r,k}^{H}\Theta^{H}\mathbf{\bar{G}}^{H}\mathbf{\tilde{G}}\Theta\bar{\Theta}}}\right.
\]

\begin{equation}
\left.+\underset{\Delta_{1,3}}{\underbrace{\sqrt{\rho_{b}}\sqrt{\rho_{k}}\mathbf{\bar{h}}_{r,k}^{H}\Theta^{H}\mathbf{\tilde{G}}^{H}\mathbf{\bar{G}}\Theta\bar{\Theta}}}+\underset{\Delta_{1,4}}{\underbrace{\sqrt{\rho_{k}}\mathbf{\bar{h}}_{r,k}^{H}\Theta^{H}\mathbf{\tilde{G}}^{H}\mathbf{\tilde{G}}\Theta\bar{\Theta}}}\right)
\end{equation}

\begin{equation}
\mathscr{E}\left\{ \left|\Delta_{1}\right|^{2}\right\} =\mathscr{E}\left\{ \left|\Delta_{1,1}\right|^{2}\right\} +\mathscr{E}\left\{ \left|\Delta_{1,2}\right|^{2}\right\} +\mathscr{E}\left\{ \left|\Delta_{1,3}\right|^{2}\right\} +\mathscr{E}\left\{ \left|\Delta_{1,4}\right|^{2}\right\} +2\mathscr{E}\left\{ \Delta_{1,1}\Delta_{1,4}^{H}\right\} 
\end{equation}

where 

\begin{equation}
\Delta_{1,1}=\rho_{b}\sqrt{\rho_{k}}\left(a_{M}^{H}\left(\phi_{kr}^{a},\phi_{kr}^{e}\right)\Theta^{H}a_{M}^{H}\left(\phi_{r}^{a},\phi_{r}^{e}\right)a_{N}^{H}\left(\phi_{b}^{a},\phi_{b}^{e}\right)a_{N}\left(\phi_{b}^{a},\phi_{b}^{e}\right)\right)\left(a_{M}\left(\phi_{r}^{a},\phi_{r}^{e}\right)\Theta\bar{\Theta}\right)
\end{equation}

\[
\mathscr{E}\left\{ \left|\Delta_{1,1}\right|^{2}\right\} =\rho_{b}^{2}\rho_{k}\left|\left(a_{M}^{H}\left(\phi_{kr}^{a},\phi_{kr}^{e}\right)\Theta^{H}a_{M}^{H}\left(\phi_{r}^{a},\phi_{r}^{e}\right)a_{N}^{H}\left(\phi_{b}^{a},\phi_{b}^{e}\right)a_{N}\left(\phi_{b}^{a},\phi_{b}^{e}\right)\right)\right|^{2}
\]

\begin{equation}
\times\mathscr{E}\left\{ \left\Vert a_{M}\left(\phi_{r}^{a},\phi_{r}^{e}\right)\Theta\bar{\Theta}\right\Vert ^{2}\right\} 
\end{equation}

\begin{equation}
\mathscr{E}\left\{ \left|\Delta_{1,1}\right|^{2}\right\} =\rho_{b}^{2}\rho_{k}\left|\left(a_{M}^{H}\left(\phi_{kr}^{a},\phi_{kr}^{e}\right)\Theta^{H}a_{M}^{H}\left(\phi_{r}^{a},\phi_{r}^{e}\right)a_{N}^{H}\left(\phi_{b}^{a},\phi_{b}^{e}\right)a_{N}\left(\phi_{b}^{a},\phi_{b}^{e}\right)\right)\right|^{2}\times M
\end{equation}

The second term can be expressed as,

\begin{equation}
\Delta_{1,2}=\sqrt{\rho_{b}}\sqrt{\rho_{k}}a_{M}^{H}\left(\phi_{kr}^{a},\phi_{kr}^{e}\right)\Theta^{H}a_{M}\left(\phi_{r}^{a},\phi_{r}^{e}\right)\stackrel[m=1]{M}{\sum}\stackrel[n=1]{N}{\sum}a_{N,n}^{H}\left(\phi_{b}^{a},\phi_{b}^{e}\right)\tilde{\mathbf{g}}_{nm}e^{j\varphi_{m}}e^{j\bar{\varphi_{m}}}
\end{equation}

\begin{equation}
\mathscr{E}\left\{ \left|\Delta_{1,2}\right|^{2}\right\} =\rho_{b}\rho_{k}\left|a_{M}^{H}\left(\phi_{kr}^{a},\phi_{kr}^{e}\right)\Theta^{H}a_{M}\left(\phi_{r}^{a},\phi_{r}^{e}\right)\right|^{2}NM
\end{equation}

The third term can be written as 

\begin{equation}
\Delta_{1,3}=\sqrt{\rho_{b}}\sqrt{\rho_{k}}\stackrel[m=1]{M}{\sum}\stackrel[n=1]{N}{\sum}a_{M,m}^{H}\left(\phi_{kr}^{a},\phi_{kr}^{e}\right)\tilde{g}_{nm}^{H}e^{-j\varphi_{m}}a_{N,n}\left(\phi_{b}^{a},\phi_{b}^{e}\right)\stackrel[m=1]{M}{\sum}a_{M,m}^{H}\left(\phi_{r}^{a},\phi_{r}^{e}\right)e^{j\bar{\varphi_{m}}}e^{j\varphi_{m}}
\end{equation}

\begin{equation}
\mathscr{E}\left\{ \left|\Delta_{1,3}\right|^{2}\right\} =\rho_{b}\rho_{k}MN\left(\mathscr{E}\left|\stackrel[m=1]{M}{\sum}a_{M,m}^{H}\left(\phi_{r}^{a},\phi_{r}^{e}\right)e^{j\bar{\varphi_{m}}}e^{j\varphi_{m}}\right|^{2}\right)
\end{equation}

\begin{equation}
\mathscr{E}\left\{ \left|\Delta_{1,3}\right|^{2}\right\} =\rho_{b}\rho_{k}MN\left(M+\rho\left(\kappa\right)^{2}\stackrel[m_{1}=1]{M}{\sum}\stackrel[m_{2}\neq m_{1}]{M}{\sum}a_{M,m_{1}}^{H}\left(\phi_{r}^{a},\phi_{r}^{e}\right)e^{j\varphi_{m_{1}}}a_{M,m_{2}}\left(\phi_{r}^{a},\phi_{r}^{e}\right)e^{-j\varphi_{m_{2}}}\right)
\end{equation}

\begin{equation}
\mathscr{E}\left\{ \left|\Delta_{1,3}\right|^{2}\right\} =\rho_{b}\rho_{k}MN\left(\rho\left(\kappa\right)^{2}M+\left(1-\rho\left(\kappa\right)^{2}\right)M\right)
\end{equation}

The forth term can be represented as,

\begin{equation}
\Delta_{1,4}=\sqrt{\rho_{k}}\stackrel[n=1]{N}{\sum}\stackrel[m_{1}=1]{M}{\sum}a_{M,m_{1}}^{H}\left(\phi_{kr}^{a},\phi_{kr}^{e}\right)e^{-j\varphi_{m}}\tilde{g}_{nm_{1}}^{H}\stackrel[m_{2}=1]{M}{\sum}\tilde{g}_{nm_{2}}e^{j\varphi_{m}}e^{j\bar{\varphi_{m}}}
\end{equation}

\begin{equation}
\mathscr{E}\left\{ \left|\Delta_{1,4}\right|^{2}\right\} =\rho_{k}\left(N^{2}M+NM^{2}\right)
\end{equation}

Now the last term can be written as 

\[
\mathscr{E}\left\{ \Delta_{1,1}\Delta_{1,4}^{*}\right\} =\rho_{b}\rho_{k}\left(a_{M}^{H}\left(\phi_{kr}^{a},\phi_{kr}^{e}\right)\Theta^{H}a_{M}^{H}\left(\phi_{r}^{a},\phi_{r}^{e}\right)a_{N}^{H}\left(\phi_{b}^{a},\phi_{b}^{e}\right)a_{N}\left(\phi_{b}^{a},\phi_{b}^{e}\right)\right)
\]

\begin{equation}
\times\left(a_{M}\left(\phi_{r}^{a},\phi_{r}^{e}\right)\Theta\bar{\Theta}\right)\rho_{k}\mathbf{\bar{h}}_{r,k}^{H}\Theta^{H}\mathbf{\tilde{G}}^{H}\mathbf{\tilde{G}}\Theta\bar{\Theta}
\end{equation}

\begin{equation}
\mathscr{E}\left\{ \Delta_{1,1}\Delta_{1,4}^{*}\right\} =\rho_{b}\rho_{k}\left(a_{M}^{H}\left(\phi_{kr}^{a},\phi_{kr}^{e}\right)\Theta^{H}a_{M}^{H}\left(\phi_{r}^{a},\phi_{r}^{e}\right)a_{N}^{H}\left(\phi_{b}^{a},\phi_{b}^{e}\right)a_{N}\left(\phi_{b}^{a},\phi_{b}^{e}\right)\right)\left(a_{M}\left(\phi_{r}^{a},\phi_{r}^{e}\right)\Theta\right)\rho_{k}\mathbf{\bar{h}}_{r,k}^{H}\Theta^{H}N\Theta
\end{equation}

We will repeat similar steps for $\Delta_{2}$,

\[
\Delta_{2}=\left(\underset{\Delta_{2,1}}{\underbrace{\rho_{b}\mathbf{\tilde{h}}_{r,k}^{H}\Theta^{H}\mathbf{\bar{G}}^{H}\mathbf{\bar{G}}\Theta\bar{\Theta}}}+\underset{\Delta_{2,2}}{\underbrace{\sqrt{\rho_{b}}\mathbf{\tilde{h}}_{r,k}^{H}\Theta^{H}\mathbf{\bar{G}}^{H}\mathbf{\tilde{G}}\Theta\bar{\Theta}}}\right.
\]

\begin{equation}
\left.+\underset{\Delta_{2,3}}{\underbrace{\sqrt{\rho_{b}}\mathbf{\tilde{h}}_{r,k}^{H}\Theta^{H}\mathbf{\tilde{G}}^{H}\mathbf{\bar{G}}\Theta\bar{\Theta}}}+\underset{\Delta_{2,4}}{\underbrace{\mathbf{\tilde{h}}_{r,k}^{H}\Theta^{H}\mathbf{\tilde{G}}^{H}\mathbf{\tilde{G}}\Theta\bar{\Theta}}}\right)
\end{equation}

\begin{equation}
\mathscr{E}\left\{ \left|\Delta_{2}\right|^{2}\right\} =\mathscr{E}\left\{ \left|\Delta_{2,1}\right|^{2}\right\} +\mathscr{E}\left\{ \left|\Delta_{2,2}\right|^{2}\right\} +\mathscr{E}\left\{ \left|\Delta_{2,3}\right|^{2}\right\} +\mathscr{E}\left\{ \left|\Delta_{2,4}\right|^{2}\right\} +2\mathscr{E}\left\{ \Delta_{2,1}\Delta_{2,4}^{H}\right\} 
\end{equation}

where 

\begin{equation}
\Delta_{2,1}=\rho_{b}\mathbf{\tilde{h}}_{r,k}^{H}\Theta^{H}\mathbf{\bar{G}}^{H}\mathbf{\bar{G}}\Theta\bar{\Theta}
\end{equation}

\begin{equation}
\mathscr{E}\left\{ \left|\Delta_{2,1}\right|^{2}\right\} =\rho_{b}^{2}\left\Vert \Theta^{H}a_{M}^{H}\left(\phi_{r}^{a},\phi_{r}^{e}\right)a_{N}^{H}\left(\phi_{b}^{a},\phi_{b}^{e}\right)a_{N}\left(\phi_{b}^{a},\phi_{b}^{e}\right)a_{M}\left(\phi_{r}^{a},\phi_{r}^{e}\right)\Theta\right\Vert _{F}^{2}
\end{equation}

and

\begin{equation}
\Delta_{2,2}=\sqrt{\rho_{b}}\mathbf{\tilde{h}}_{r,k}^{H}\Theta^{H}a_{M}\left(\phi_{r}^{a},\phi_{r}^{e}\right)\stackrel[m=1]{M}{\sum}\stackrel[n=1]{N}{\sum}a_{N,n}^{H}\left(\phi_{b}^{a},\phi_{b}^{e}\right)\tilde{\mathbf{g}}_{nm}e^{j\varphi_{m}}e^{j\bar{\varphi_{m}}}
\end{equation}

\begin{equation}
\mathscr{E}\left\{ \left|\Delta_{2,2}\right|^{2}\right\} =\rho_{b}\left|\mathbf{\tilde{h}}_{r,k}^{H}\Theta^{H}a_{M}\left(\phi_{r}^{a},\phi_{r}^{e}\right)\right|^{2}NM
\end{equation}

while

\begin{equation}
\Delta_{2,3}=\sqrt{\rho_{b}}\stackrel[m=1]{M}{\sum}\stackrel[n=1]{N}{\sum}\tilde{h}_{r,k,nm}^{H}\tilde{g}_{nm}^{H}e^{-j\varphi_{m}}a_{N,n}\left(\phi_{b}^{a},\phi_{b}^{e}\right)\stackrel[m=1]{M}{\sum}a_{M,m}^{H}\left(\phi_{r}^{a},\phi_{r}^{e}\right)e^{j\bar{\varphi_{m}}}e^{j\varphi_{m}}
\end{equation}

\begin{equation}
\mathscr{E}\left\{ \left|\Delta_{2,3}\right|^{2}\right\} =\rho_{b}MN\left(\mathscr{E}\left|\stackrel[m=1]{M}{\sum}a_{M,m}^{H}\left(\phi_{r}^{a},\phi_{r}^{e}\right)e^{j\bar{\varphi_{m}}}e^{j\varphi_{m}}\right|^{2}\right)=\rho_{b}MN\left(\rho\left(\kappa\right)^{2}M+\left(1-\rho\left(\kappa\right)^{2}\right)M\right)
\end{equation}

Finally,

\begin{equation}
\Delta_{2,4}=\rho_{k}\stackrel[n=1]{N}{\sum}\stackrel[m_{1}=1]{M}{\sum}\tilde{h}_{r,k,nm_{1}}^{H}e^{-j\varphi_{m}}\tilde{g}_{nm_{1}}^{H}\stackrel[m_{2}=1]{M}{\sum}\tilde{g}_{nm_{2}}e^{j\varphi_{m}}e^{j\bar{\varphi_{m}}}
\end{equation}

\begin{equation}
\mathscr{E}\left\{ \left|\Delta_{2,4}\right|^{2}\right\} =\rho_{k}^{2}\left(N^{2}M+NM^{2}\right)
\end{equation}

and

\[
\mathscr{E}\left\{ \Delta_{2,1}\Delta_{2,4}^{*}\right\} =\mathscr{E}\left\{ \rho_{b}\rho_{k}\left(\mathbf{\tilde{h}}_{r,k}^{H}\Theta^{H}a_{M}^{H}\left(\phi_{r}^{a},\phi_{r}^{e}\right)a_{N}^{H}\left(\phi_{b}^{a},\phi_{b}^{e}\right)a_{N}\left(\phi_{b}^{a},\phi_{b}^{e}\right)\right)\left(a_{M}\left(\phi_{r}^{a},\phi_{r}^{e}\right)\Theta\bar{\Theta}\right)\rho_{k}\mathbf{\tilde{h}}_{r,k}^{H}\Theta^{H}\mathbf{\tilde{G}}^{H}\mathbf{\tilde{G}}\Theta\bar{\Theta}\right\} 
\]

\begin{equation}
=\rho_{k}\left(\Theta^{H}a_{M}^{H}\left(\phi_{r}^{a},\phi_{r}^{e}\right)a_{N}^{H}\left(\phi_{b}^{a},\phi_{b}^{e}\right)a_{N}\left(\phi_{b}^{a},\phi_{b}^{e}\right)\right)\left(a_{M}\left(\phi_{r}^{a},\phi_{r}^{e}\right)\Theta\right)\rho_{k}\Theta N\Theta^{H}
\end{equation}

\bibliographystyle{IEEEtran}
\bibliography{Secrecy_RIS}

\end{document}